\PassOptionsToPackage{unicode}{hyperref}
\PassOptionsToPackage{hyphens}{url}
\PassOptionsToPackage{dvipsnames,svgnames,x11names}{xcolor}
\documentclass[
  letterpaper,
  DIV=11,
  numbers=noendperiod]{scrartcl}

\usepackage{amsmath,amssymb}
\usepackage{iftex}
\ifPDFTeX
  \usepackage[T1]{fontenc}
  \usepackage[utf8]{inputenc}
  \usepackage{textcomp} 
\else 
  \usepackage{unicode-math}
  \defaultfontfeatures{Scale=MatchLowercase}
  \defaultfontfeatures[\rmfamily]{Ligatures=TeX,Scale=1}
\fi
\usepackage{lmodern}
\ifPDFTeX\else  
\fi
\IfFileExists{upquote.sty}{\usepackage{upquote}}{}
\IfFileExists{microtype.sty}{
  \usepackage[]{microtype}
  \UseMicrotypeSet[protrusion]{basicmath} 
}{}
\makeatletter
\@ifundefined{KOMAClassName}{
  \IfFileExists{parskip.sty}{%
    \usepackage{parskip}
  }{
    \setlength{\parindent}{0pt}
    \setlength{\parskip}{6pt plus 2pt minus 1pt}}
}{
  \KOMAoptions{parskip=half}}
\makeatother
\usepackage{xcolor}
\ifx\paragraph\undefined\else
  \let\oldparagraph\paragraph
  \renewcommand{\paragraph}[1]{\oldparagraph{#1}\mbox{}}
\fi
\ifx\subparagraph\undefined\else
  \let\oldsubparagraph\subparagraph
  \renewcommand{\subparagraph}[1]{\oldsubparagraph{#1}\mbox{}}
\fi

\providecommand{\tightlist}{%
  \setlength{\itemsep}{0pt}\setlength{\parskip}{0pt}}\usepackage{longtable,booktabs,array}
\usepackage{calc} 
\usepackage{etoolbox}
\makeatletter
\patchcmd\longtable{\par}{\if@noskipsec\mbox{}\fi\par}{}{}
\makeatother
\IfFileExists{footnotehyper.sty}{\usepackage{footnotehyper}}{\usepackage{footnote}}
\makesavenoteenv{longtable}
\usepackage{graphicx}
\makeatletter
\def\maxwidth{\ifdim\Gin@nat@width>\linewidth\linewidth\else\Gin@nat@width\fi}
\def\maxheight{\ifdim\Gin@nat@height>\textheight\textheight\else\Gin@nat@height\fi}
\makeatother
\setkeys{Gin}{width=\maxwidth,height=\maxheight,keepaspectratio}
\makeatletter
\def\fps@figure{htbp}
\makeatother
\NewDocumentCommand\citeproctext{}{}

\makeatletter
 \let\@cite@ofmt\@firstofone
 \def\@biblabel#1{}
 \def\@cite#1#2{{#1\if@tempswa , #2\fi}}
\makeatother
\newlength{\cslhangindent}
\newlength{\csllabelwidth}
\newenvironment{CSLReferences}[2] 
 {\begin{list}{}{%
  \setlength{\itemindent}{0pt}
  \setlength{\leftmargin}{0pt}
  \setlength{\parsep}{0pt}
  \ifodd #1
   \setlength{\leftmargin}{\cslhangindent}
   \setlength{\itemindent}{-1\cslhangindent}
  \fi
  \setlength{\itemsep}{#2\baselineskip}}}
 {\end{list}}
\usepackage{calc}

\KOMAoption{captions}{tableheading}
\makeatletter
\@ifpackageloaded{caption}{}{\usepackage{caption}}
\AtBeginDocument{%
\ifdefined\contentsname
  \renewcommand*\contentsname{Table of contents}
\else
  \newcommand\contentsname{Table of contents}
\fi
\ifdefined\listfigurename
  \renewcommand*\listfigurename{List of Figures}
\else
  \newcommand\listfigurename{List of Figures}
\fi
\ifdefined\listtablename
  \renewcommand*\listtablename{List of Tables}
\else
  \newcommand\listtablename{List of Tables}
\fi
\ifdefined\figurename
  \renewcommand*\figurename{Figure}
\else
  \newcommand\figurename{Figure}
\fi
\ifdefined\tablename
  \renewcommand*\tablename{Table}
\else
  \newcommand\tablename{Table}
\fi
}
\@ifpackageloaded{float}{}{\usepackage{float}}
\floatstyle{ruled}
\@ifundefined{c@chapter}{\newfloat{codelisting}{h}{lop}}{\newfloat{codelisting}{h}{lop}[chapter]}
\floatname{codelisting}{Listing}

\makeatother
\makeatletter
\@ifpackageloaded{caption}{}{\usepackage{caption}}
\@ifpackageloaded{subcaption}{}{\usepackage{subcaption}}
\makeatother
\ifLuaTeX
  \usepackage{selnolig}  
\fi
\usepackage{bookmark}

\IfFileExists{xurl.sty}{\usepackage{xurl}}{} 
\hypersetup{
  pdftitle={Time-lapse full-waveform permeability inversion: a feasibility study},
  pdfauthor={Ziyi Yin1, Mathias Louboutin1,3, Olav Møyner2, Felix J. Herrmann1 1 Georgia Institute of Technology, 2 SINTEF Digital, 3 Now at Devito Codes Ltd},
  colorlinks=true,
  linkcolor={blue},
  filecolor={Maroon},
  citecolor={Blue},
  urlcolor={Blue},
  pdfcreator={LaTeX via pandoc}}

\title{Time-lapse full-waveform permeability inversion: a feasibility
study}
\author{Ziyi Yin\textsuperscript{1}, Mathias
Louboutin\textsuperscript{1,3}, Olav Møyner\textsuperscript{2}, Felix J.
Herrmann\textsuperscript{1}\\
\textsuperscript{1} Georgia Institute of Technology, \textsuperscript{2}
SINTEF Digital, \textsuperscript{3} Now at Devito Codes Ltd}
\date{}

\begin{document}
\maketitle

\section{Abstract}\label{abstract}

Time-lapse seismic monitoring necessitates integrated workflows that
combine seismic and reservoir modeling to enhance reservoir property
estimation. We present a feasibility study of an end-to-end inversion
framework that directly inverts for permeability from prestack
time-lapse seismic data. To assess the method's robustness, we design
experiments focusing on its sensitivity to initial models and potential
errors in modeling. Our study leverages the Compass model to simulate
CO\textsubscript{2} storage in saline aquifers, which is derived from
well and seismic data from the North Sea, a candidate site for
geological carbon storage.

\section{Introduction}\label{introduction}

Despite significant advancements in reservoir monitoring over recent
decades, time-lapse seismic technology continues to grapple with
challenges related to cost and efficiency (D. E. Lumley 2001; R. A.
Chadwick et al. 2009; A. Chadwick et al. 2010; Furre et al. 2017).
Employing 4D seismic workflows, including time-lapse full-waveform
inversion (TL-FWI, D. Lumley 2010; Hicks et al. 2016), has become a
common practice for estimating changes in the Earth's elastic
properties, facilitating the quantitative interpretation of these
changes as indicators of reservoir attributes like fluid content and
pressure (Bosch, Mukerji, and Gonzalez 2010; Wei et al. 2017). Recent
methodologies aim to leverage time-lapse seismic data for the joint
estimation of both elastic and reservoir properties, with a focus on
parameters such as saturation and porosity (Bosch et al. 2007; Hu,
Grana, and Innanen 2022). However, the integration of seismic imaging
workflows with reservoir simulation tools remains limited, constraining
the direct application of time-lapse seismic data for permeability
estimation or updates. While a few methodologies, such as the approach
in Eikrem et al. (2016), offer some integration by using ensemble Kalman
filtering to refine permeability and porosity estimates, and Don W.
Vasco et al. (2004); D. W. Vasco et al. (2008) have explored using
time-lapse seismic data for linearized inversion to update permeability,
these efforts represent initial steps toward a more integrated approach.

This paper introduces a novel 4D processing framework for estimating
permeability directly from prestack time-lapse seismic data, offering a
streamlined, geophysically based inversion process. Unlike traditional
methods, this framework, tested in various synthetic case studies (Li et
al. 2020; Yin et al. 2022; Louboutin, Yin, et al. 2023; Yin, Orozco, et
al. 2023b), updates permeability models by exclusively matching observed
time-lapse seismic data. Despite the potential for rapid model updates,
initial results have not yet demonstrated significant alterations in
fluid saturation predictions, and the resulting permeability models
often lack the heterogeneity necessary for detailed analysis. To address
these limitations and assess the framework's real-world applicability
for 4D monitoring, we undertake a feasibility study using the Compass
model (E. Jones et al. 2012), chosen for its relevance to the North
Sea's geological structures --- a region under consideration for
CO\textsubscript{2} storage. This study evaluates the framework's
sensitivity to different starting models and forward modeling errors,
omitting regularization techniques to focus on the impact of seismic
data on permeability updates. We explore the framework's ability to
recover fine permeability structures, predict CO\textsubscript{2}
dynamics within the seismic monitoring period, and forecast
CO\textsubscript{2} dynamics in near future without any seismic
observation. Recognizing the limitations of our simplifying assumptions,
we conclude with suggestions for future research to advance this
promising field.

\section{Permeability inversion
framework}\label{permeability-inversion-framework}

\begin{figure}

\centering{

\includegraphics{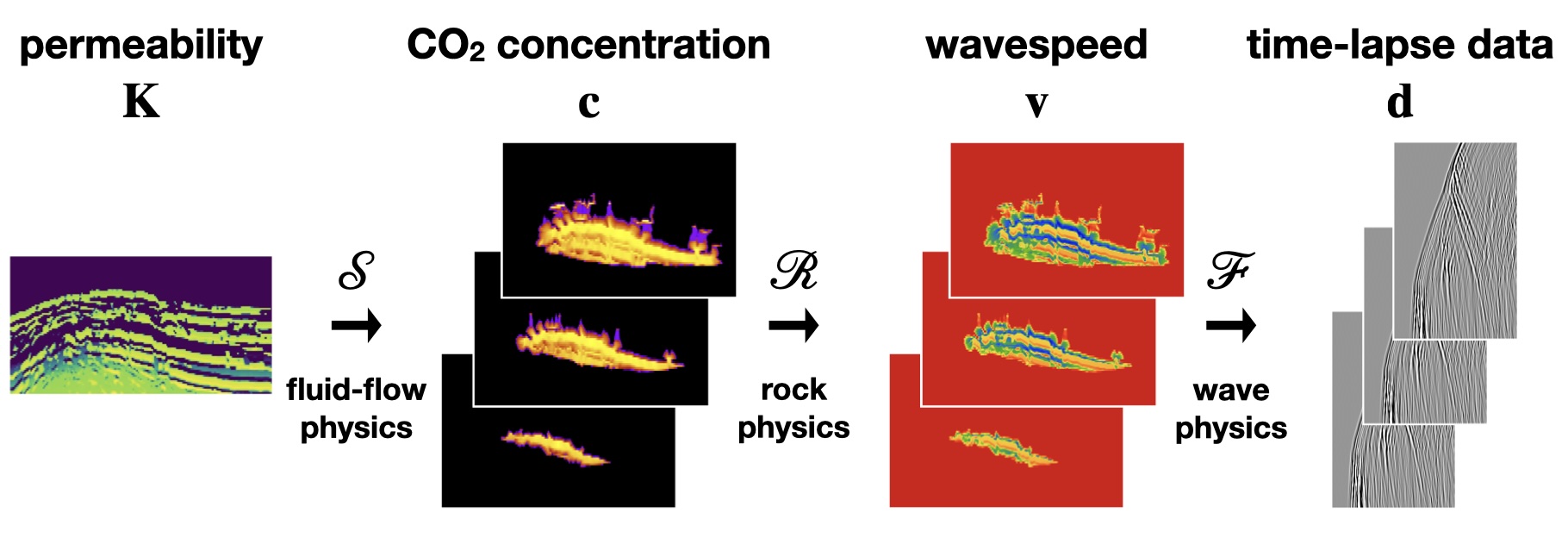}

}

\caption{\label{fig-multiphysics}Multiphysics forward model.}

\end{figure}

Our feasibility study examines the time-lapse seismic monitoring of
CO\textsubscript{2} storage, focusing on the integration of three
fundamental physics disciplines critical to this process: fluid-flow
physics, rock physics, and wave physics, as illustrated in
Figure~\ref{fig-multiphysics}. The dynamics of the CO\textsubscript{2}
plume during injection are modeled using multiphase flow equations
(Pruess and Nordbotten 2011), processed through a reservoir simulator
(Krogstad et al. 2015; Settgast et al. 2018; Rasmussen et al. 2021;
Stacey and Williams 2017). These simulations require detailed inputs,
including well operation parameters and the spatial distribution of
porosity and permeability, while we particularly focus on permeability,
\(\mathbf{K}\), as the parameter of interest in this exposition. The
output from the reservoir simulator, \(\mathcal{S}\), primarily the
time-evolving fluid saturation of the CO\textsubscript{2} plume,
\(\mathbf{c}\), serves as the input to the rock physics model,
\(\mathcal{R}\). This model translates changes in fluid saturation
within the reservoir rocks into variations in compressional wavespeeds,
\(\mathbf{v}\), over time (Avseth, Mukerji, and Mavko 2010). Lastly,
based on the wavespeed models for each snapshot, the wave modeling
operator (Tarantola 1984), \(\mathcal{F}\), is used to generate the
time-lapse seismic data, \(\mathbf{d}\), which compiles the seismic data
from each vintage.

In practice, the prestack time-lapse seismic dataset, \(\mathbf{d}\), is
observed from the field, with the objective being to estimate the past,
current, and future dynamics of the CO\textsubscript{2} plume in GCS
projects. Our methodology diverges from traditional workflows that
typically proceed from seismic inversion to quantitative interpretation
and subsequent reservoir parameter updates based on the derived wave
properties. Instead, we propose an integrated, end-to-end approach that
directly inverts for permeability, \(\mathbf{K}\), to reduce the
time-lapse seismic data misfit. The core of our method is the
composition of three physics-based modeling operators, formulated to
minimize the following objective function:

\[\underset{\mathbf{K}}{\operatorname{minimize}} \quad \|\mathcal{F}\circ\mathcal{R}\circ\mathcal{S}(\mathbf{K})-\mathbf{d}\|_2^2.\]

This objective is minimized via an iterative process that includes:

\begin{itemize}
\tightlist
\item
  Generating synthetic time-lapse seismic data using an initial guess
  for the permeability model;
\item
  Calculating the gradient of the permeability by backpropagating the
  residuals of the seismic data;
\item
  Updating the permeability model to reduce the misfit between the
  synthetic and observed time-lapse seismic data.
\end{itemize}

The advantage of this end-to-end inversion framework lies in its ability
to break down silos through multiphysics integration. Specifically, it
eliminates the need for intermediate processing steps to update the
saturation and wavespeed models. As we demonstrate in the subsequent
feasibility study, the inverted permeability can produce more accurate
fluid saturation and wavespeed models, leading to a better fit to the
observed time-lapse seismic data.

\section{Feasibility study on the Compass
model}\label{feasibility-study-on-the-compass-model}

We evaluates the performance of this inversion framework through a
synthetic case study on the Compass model (E. Jones et al. 2012). This
model was was derived from well logs and imaged seismic from the North
Sea area, which is currently in consideration for GCS projects. We
specifically choose this model because it is representative for 4D
monitoring of GCS, and also because the grid spacing in this model is
\(6\mathrm{m}\) in both horizontal and vertical directions. Conducting a
case study in such a fine discretization can demonstrate the efficacy of
the end-to-end inversion framework for inverting fine-scale geological
structures in the permeability model.

\begin{figure}

\begin{minipage}{\linewidth}

\centering{

\captionsetup{labelsep=none}\includegraphics[width=0.96\textwidth,height=\textheight]{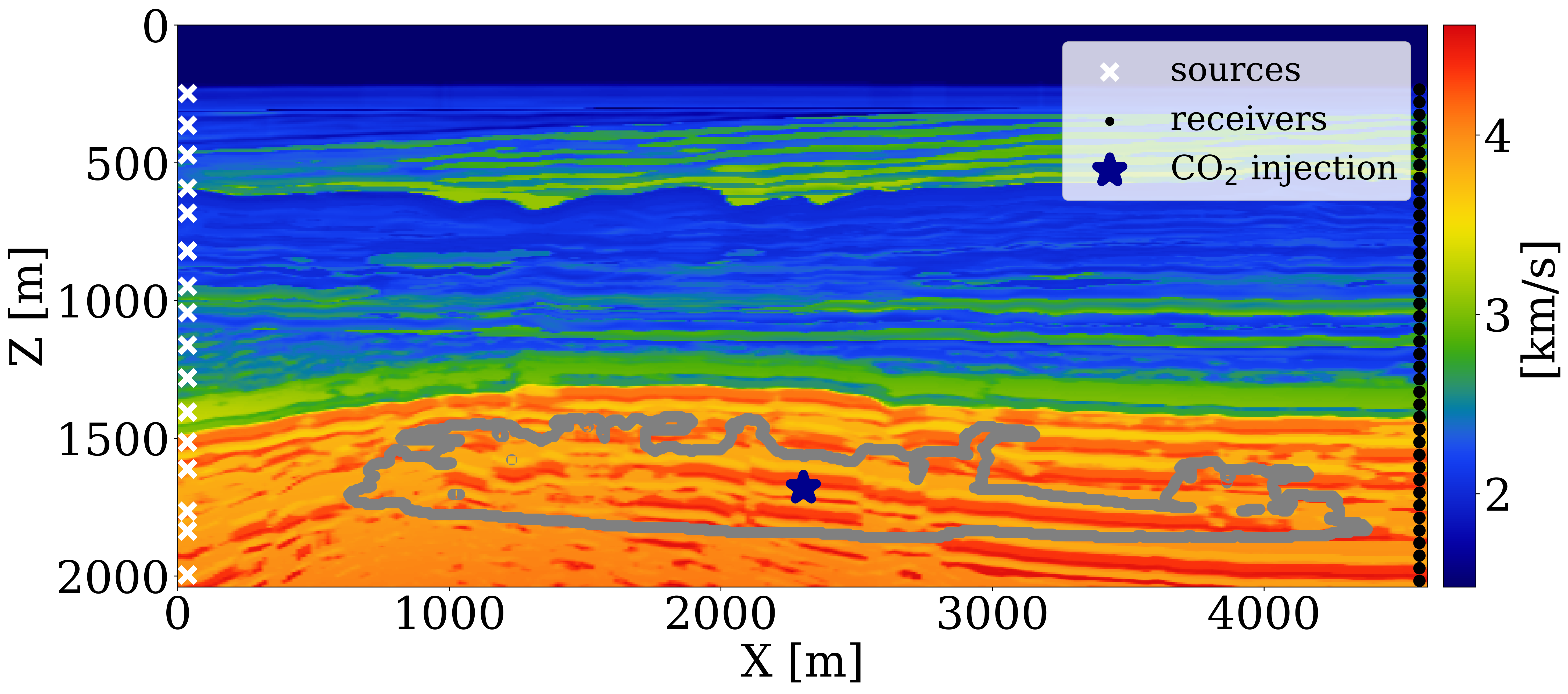}

}

\subcaption{\label{fig-acquisition}}

\end{minipage}%
\newline
\begin{minipage}{\linewidth}

\centering{

\captionsetup{labelsep=none}\includegraphics{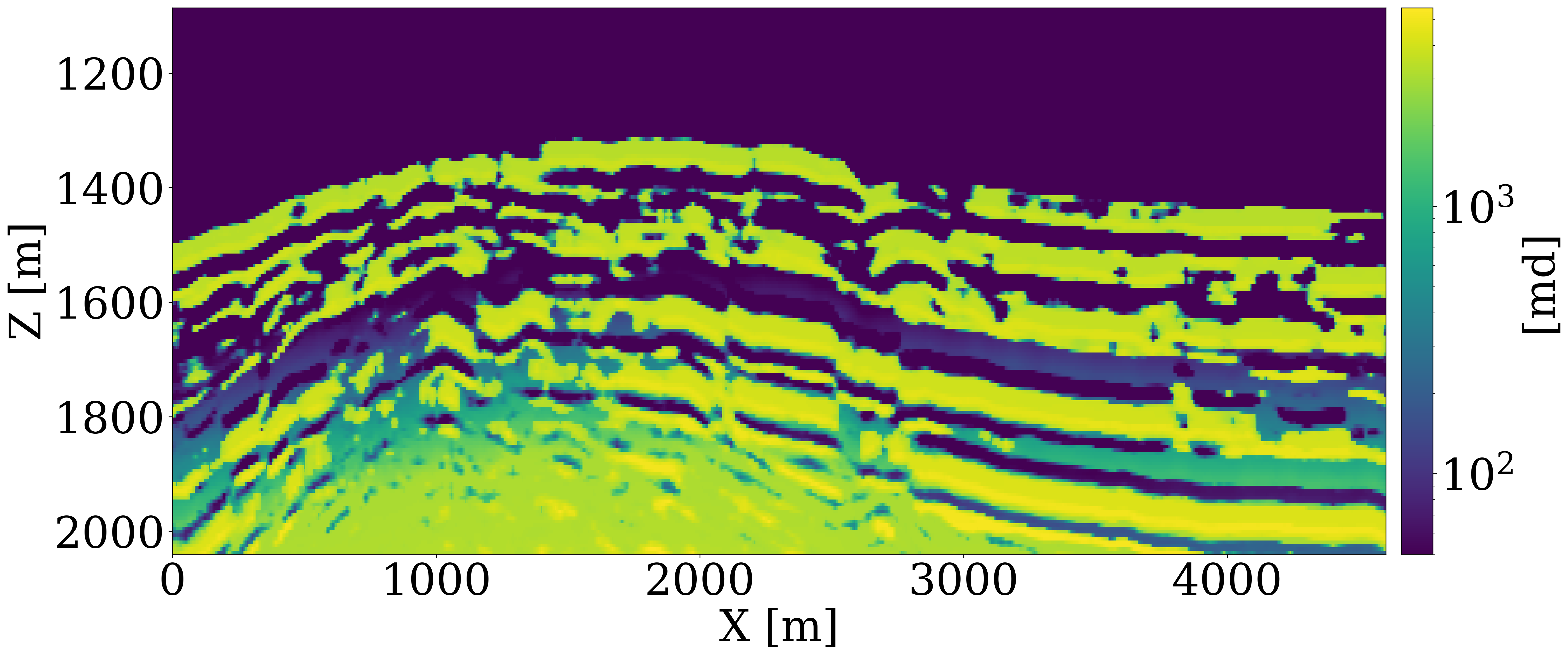}

}

\subcaption{\label{fig-true-perm}}

\end{minipage}%

\caption{\label{fig-perms}Experimental configuration. (a) Setup of
seismic acquisition and well control. Dark blue \(\star\) denotes the
CO\textsubscript{2} injection well. While \(\times\) and black \(\cdot\)
represent source and receiver locations, respectively. The gray curve
delineates the shape of the CO\textsubscript{2} plume at the 25th year.
(b) Ground truth spatial distribution of horizontal permeability.}

\end{figure}

We aim to invert for the spatial distribution of permeability using five
vintages of prestack time-lapse seismic data. We utilize a 2D slice of
the velocity model, as illustrated in Figure~\ref{fig-acquisition},
where the red region signifies the storage complex. An artificial proxy
permeability model is created to ensure significant permeability
contrasts within different layers of the storage complex, with the
horizontal permeability displayed in Figure~\ref{fig-true-perm}. The low
permeability layers range from approximately \(10^{-3}\) to \(1\)
millidarcies (md), while the high permeability layers vary between
\(600\) to \(6000\) md. A CO\textsubscript{2} injection well, marked
with black \(\times\), is placed centrally to inject supercritical
CO\textsubscript{2} for 25 years at a constant rate of 2 million metric
tons per year. We assume the porosity and the kv/kh ratio are
homogeneous and known during inversion, of values 0.25 and 10\%,
respectively. The simulation of compressible and immiscible two-phase
flow, where CO\textsubscript{2} displaces brine in porous rocks, is
performed using a fully implicit method implemented in
\href{https://github.com/sintefmath/JutulDarcy.jl}{JutulDarcy.jl}
(Møyner and Bruer 2023; Ziyi Yin, Bruer, and Louboutin 2023). The
CO\textsubscript{2} plume at the 25th year is depicted in grey in
Figure~\ref{fig-acquisition}. After converting CO\textsubscript{2}
saturation into wavespeed models via the patchy saturation model, we
generate acoustic time-lapse seismic data for five vintages at years 5,
10, 15, 20, and 25 using
\href{https://github.com/devitocodes/devito}{Devito} (Louboutin et al.
2019; Luporini et al. 2020) and
\href{https://github.com/slimgroup/JUDI.jl}{JUDI.jl} (Witte et al. 2019;
Louboutin, Witte, et al. 2023), employing a Ricker wavelet with a
central frequency of 20 Hz. The source and receiver geometries are shown
in Figure~\ref{fig-acquisition}.

\subsection{Sensitivity with respect to starting
models}\label{sensitivity-with-respect-to-starting-models}

\begin{figure}

\begin{minipage}{0.50\linewidth}

\centering{

\captionsetup{labelsep=none}\includegraphics{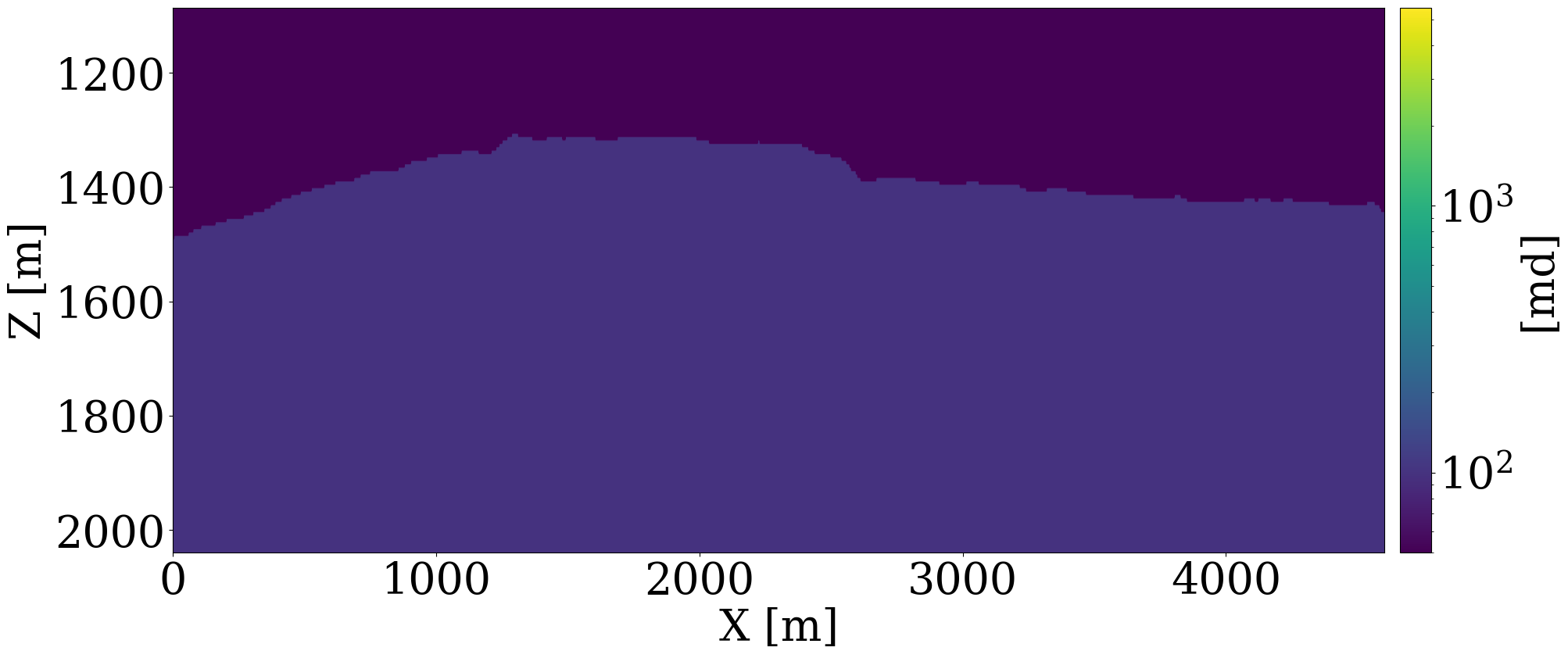}

}

\subcaption{\label{fig-init-perm-1}}

\end{minipage}%
\begin{minipage}{0.50\linewidth}

\centering{

\captionsetup{labelsep=none}\includegraphics{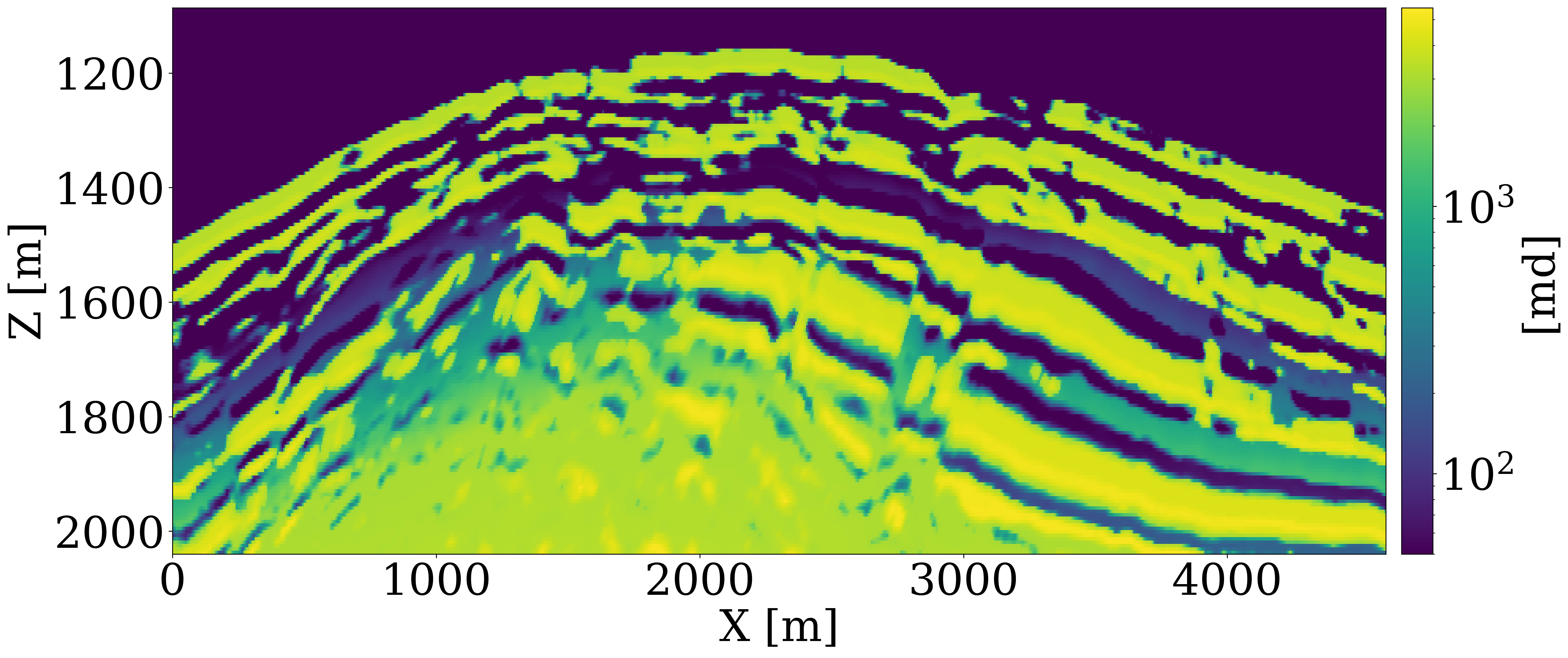}

}

\subcaption{\label{fig-init-perm-2}}

\end{minipage}%
\newline
\begin{minipage}{0.50\linewidth}

\centering{

\captionsetup{labelsep=none}\includegraphics{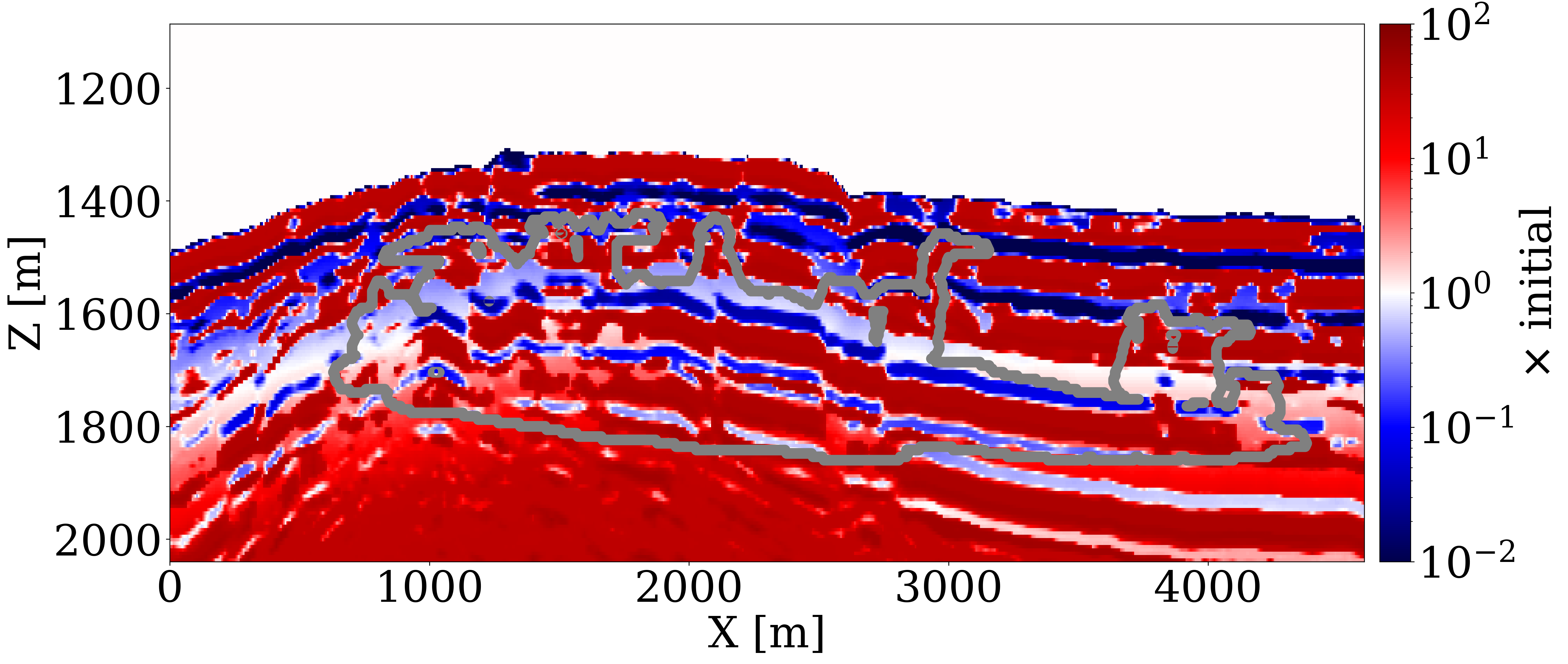}

}

\subcaption{\label{fig-case1-ideal-update}}

\end{minipage}%
\begin{minipage}{0.50\linewidth}

\centering{

\captionsetup{labelsep=none}\includegraphics{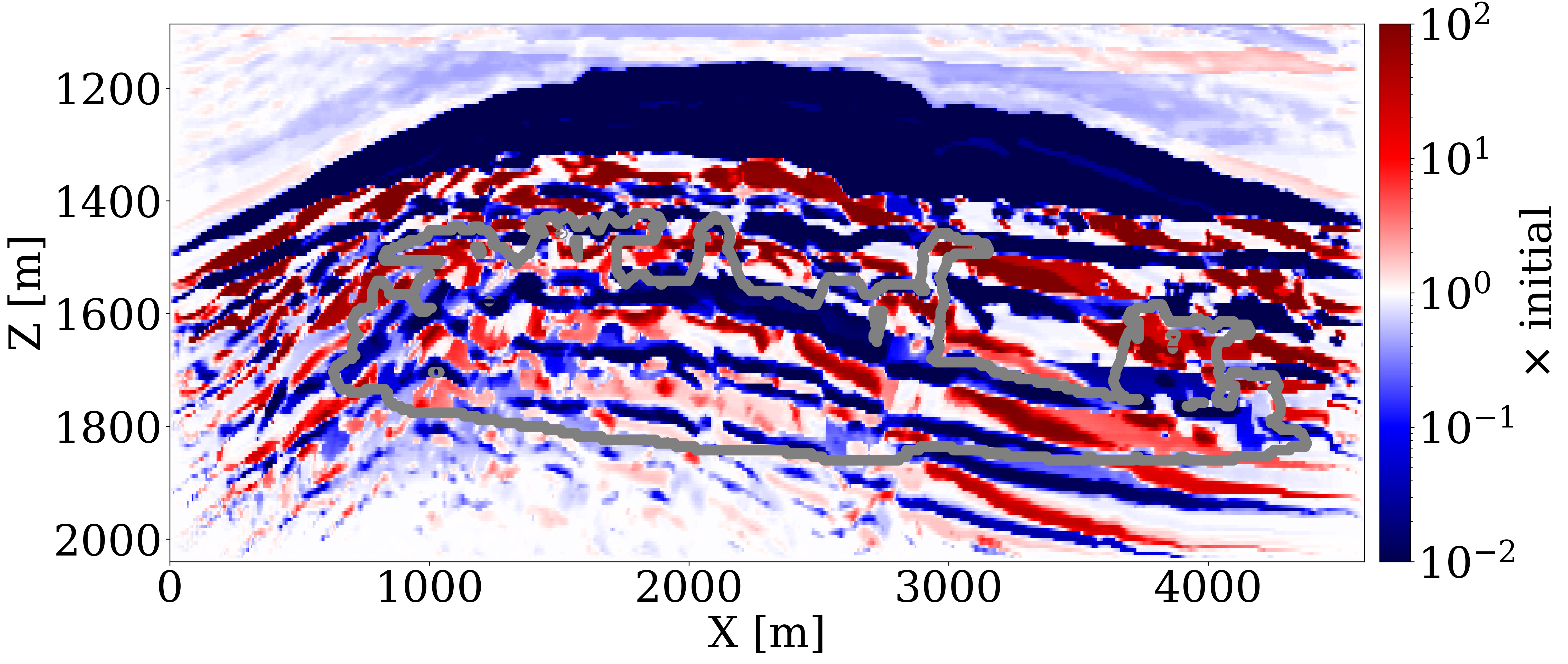}

}

\subcaption{\label{fig-case2-ideal-update}}

\end{minipage}%
\newline
\begin{minipage}{0.50\linewidth}

\centering{

\captionsetup{labelsep=none}\includegraphics{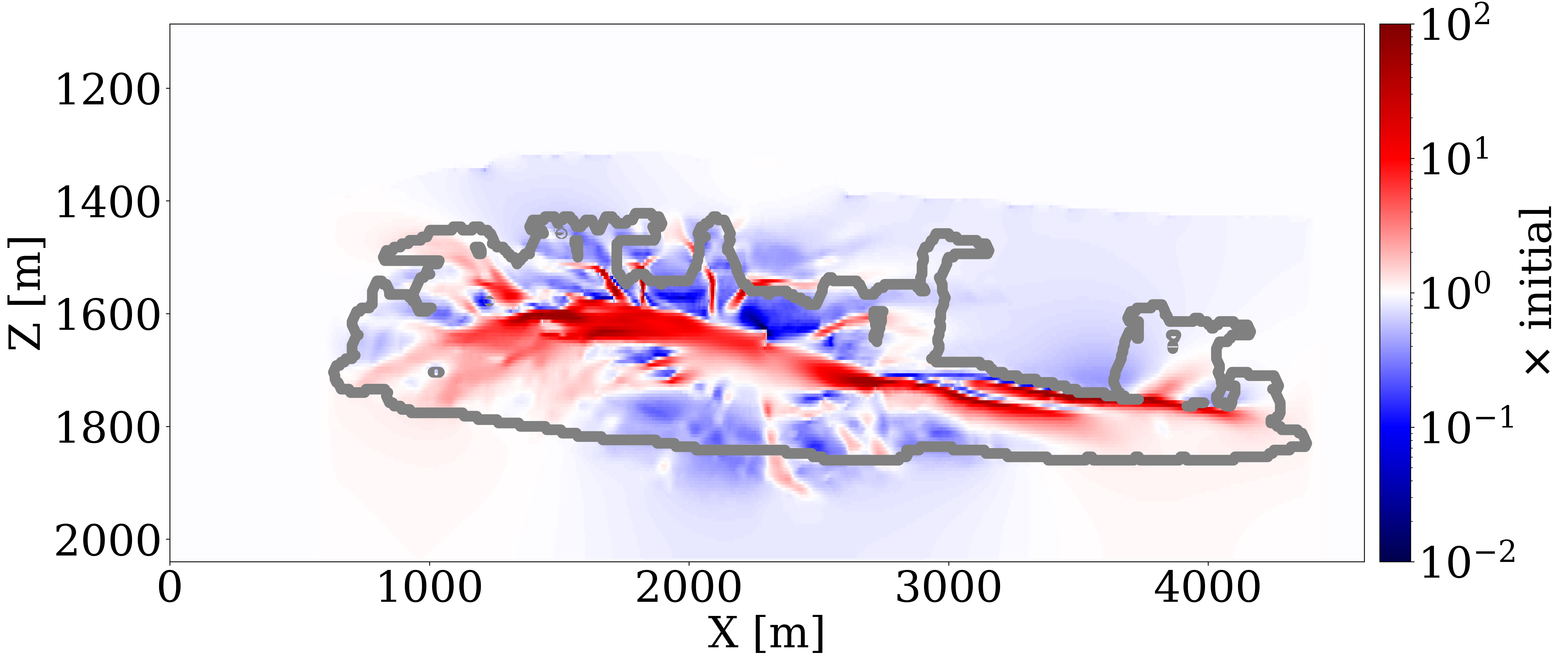}

}

\subcaption{\label{fig-case1-update-perm}}

\end{minipage}%
\begin{minipage}{0.50\linewidth}

\centering{

\captionsetup{labelsep=none}\includegraphics{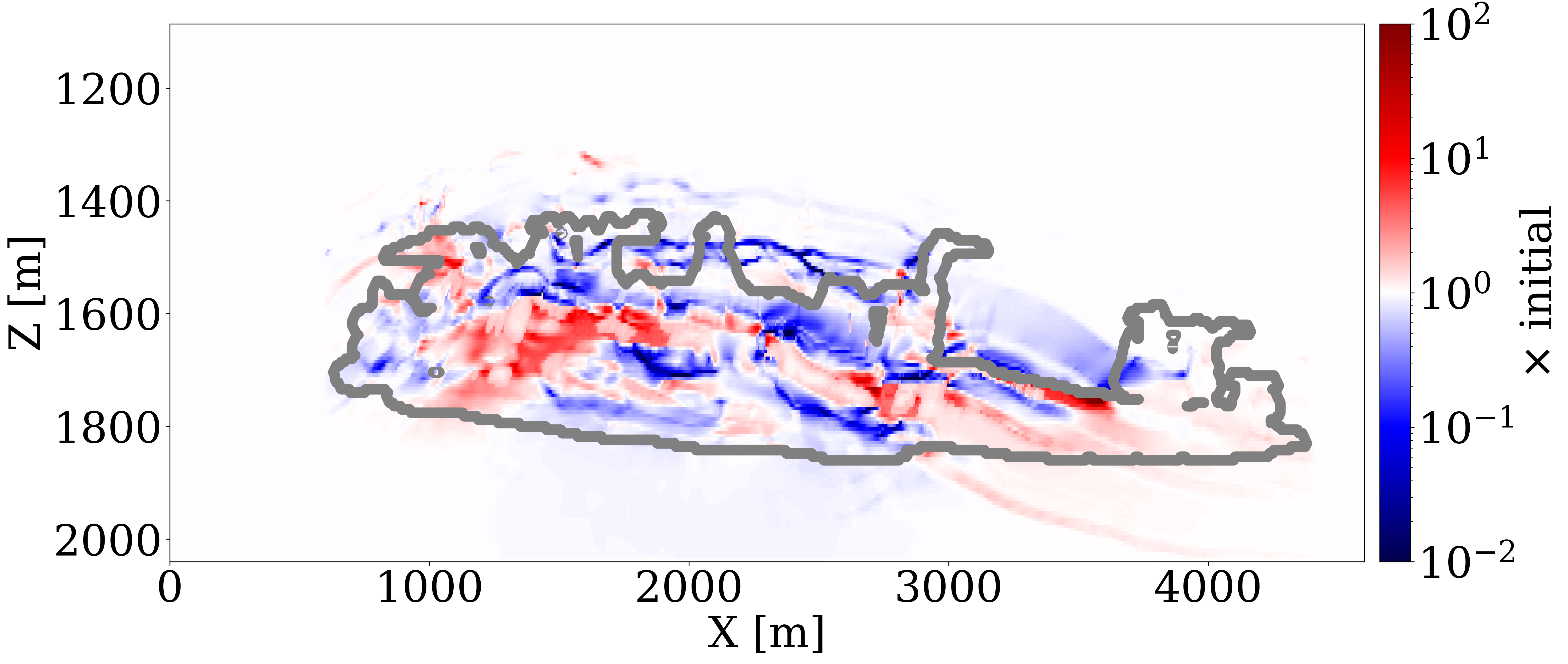}

}

\subcaption{\label{fig-case2-update-perm}}

\end{minipage}%

\caption{\label{fig-inv-perm}Permeability inversion results. (a)(c)(e)
display the initial permeability model in case 1, the logarithmic ratio
of the ground truth permeability (Figure~\ref{fig-true-perm}) to the
initial one, and the logarithmic ratio of the inverted permeability to
the initial one. (b)(d)(f) display the same but for case 2 with a
distorted initial permeability model. Gray curve indicates the boundary
of the CO\textsubscript{2} plume at the 25th year.}

\end{figure}

To evaluate the efficacy of our end-to-end inversion framework,
particularly its sensitivity to initial permeability models, we examine
two distinct starting permeability models shown in
Figure~\ref{fig-init-perm-1} and Figure~\ref{fig-init-perm-2}. The first
initial model features homogeneous permeability values
(\(100\mathrm{md}\)) across the entire reservoir. This simplistic model
serves as a blank canvas, allowing us to explore the extent of
permeability updates achievable from a non-informative permeability
model. The second scenario introduces a more complex initial model with
accurately valued permeability layers that are, however, misplaced due
to a distortion, reflecting potential real-world inaccuracies in layer
positioning (Bloice, Stocker, and Holzinger 2017). This setup aims to
simulate the challenges faced when initial rock type interpretations are
incorrect prior to CO\textsubscript{2} injection.

In both scenarios, we employ a methodological shortcut often referred to
as committing an ``inversion crime'', where the data generation and
inversion processes share the same computational kernel. This ideal
setup is used here to show what is ideally achievable by this inversion
framework, with the spatial distribution of horizontal permeability
being the sole unknown in our study. To add a layer of realism, we
incorporate 8 dB of incoherent band-limited Gaussian noise into the
observed data, mirroring the kind of data quality challenges encountered
in real seismic datasets.

After 12.5 datapass through the entire time-lapse seismic dataset, we
achieve updates in the permeability models, presented in logarithmic
scale for both initial scenarios (Figure~\ref{fig-case1-update-perm} and
Figure~\ref{fig-case2-update-perm}). Additionally,
Figure~\ref{fig-case1-ideal-update} and
Figure~\ref{fig-case2-ideal-update} offer a visualization of ``ideal''
updates by showing the logarithmic differences between the ground truth
and the initial permeability models.

The following observations are made: First, the permeability updates are
primarily confined to areas directly influenced by the
CO\textsubscript{2} plume, as delineated by gray curves. This outcome
aligns with expectations since the time-lapse variations in wave
properties are attributed exclusively to changes in fluid saturation.
Consequently, without additional information, this inversion method does
not alter permeability values outside the CO\textsubscript{2} plume's
extent. Second, the inverted permeability within the CO\textsubscript{2}
plume largely reflects the trend of the ground truth permeability model.
For the scenario with a uniform initial permeability (Case 1), the
framework successfully identifies major permeability variations --- both
high and low (depicted in red and blue, respectively) at approximately
\(1600\,\mathrm{m}\) depth --- accurately capturing their depth and
lateral distribution in alignment with the actual layers. In the
scenario involving a distorted initial model (Case 2), the inversion
process introduces high-resolution details to the layers affected by the
plume, aligning well with the ideal updates shown in
Figure~\ref{fig-case2-ideal-update}. Despite these successes, the full
magnitude of permeability contrasts is not entirely captured, pointing
to the inherently ill-posed nature of permeability inversion (Zhang,
Jafarpour, and Li 2014). This limitation highlights the need for
comprehensive uncertainty quantification (Gahlot et al. 2023) in future
studies to refine the accuracy and reliability of permeability models
derived from seismic data.

\subsection{Sensitivity with respect to forward modeling
errors}\label{sensitivity-with-respect-to-forward-modeling-errors}

To extend our investigation beyond the overly idealized scenarios, we
next examine the framework's robustness in scenarios that avoid
``committing the inversion crime''. Specifically, this means that the
forward modeling kernel used during the inversion process differs from
the one used to generate the observed data. A critical area of focus is
the potential discrepancies in the brine-filled baseline velocity model
before CO\textsubscript{2} injection --- a common source of error in
time-lapse seismic studies. Errors in this baseline model, which feeds
into the rock physics model, can produce inaccurate estimated velocity
models for CO\textsubscript{2}-filled reservoir at various stages,
leading to inaccuracies in the simulated time-lapse seismic data.

\begin{figure}

\begin{minipage}{0.50\linewidth}

\centering{

\captionsetup{labelsep=none}\includegraphics{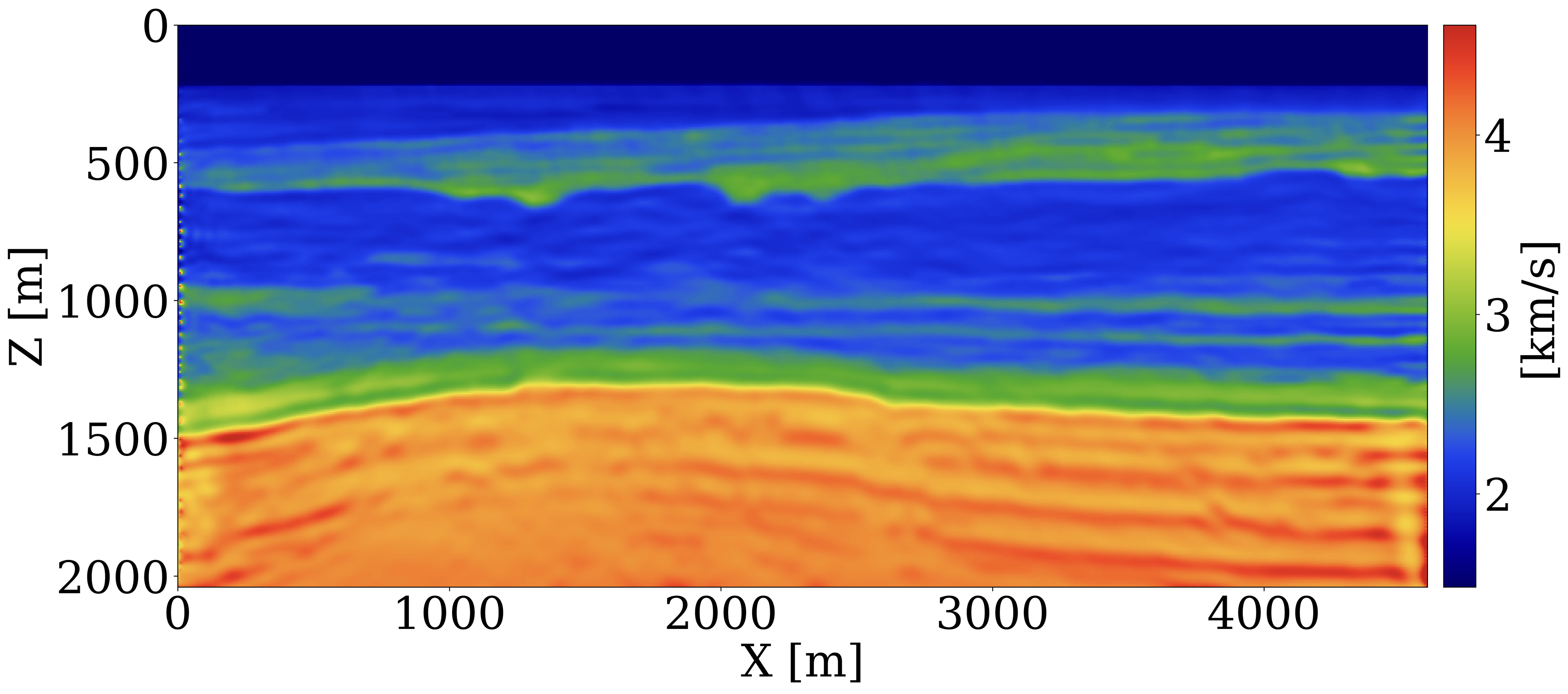}

}

\subcaption{\label{fig-fwi-v}}

\end{minipage}%
\begin{minipage}{0.50\linewidth}

\centering{

\captionsetup{labelsep=none}\includegraphics{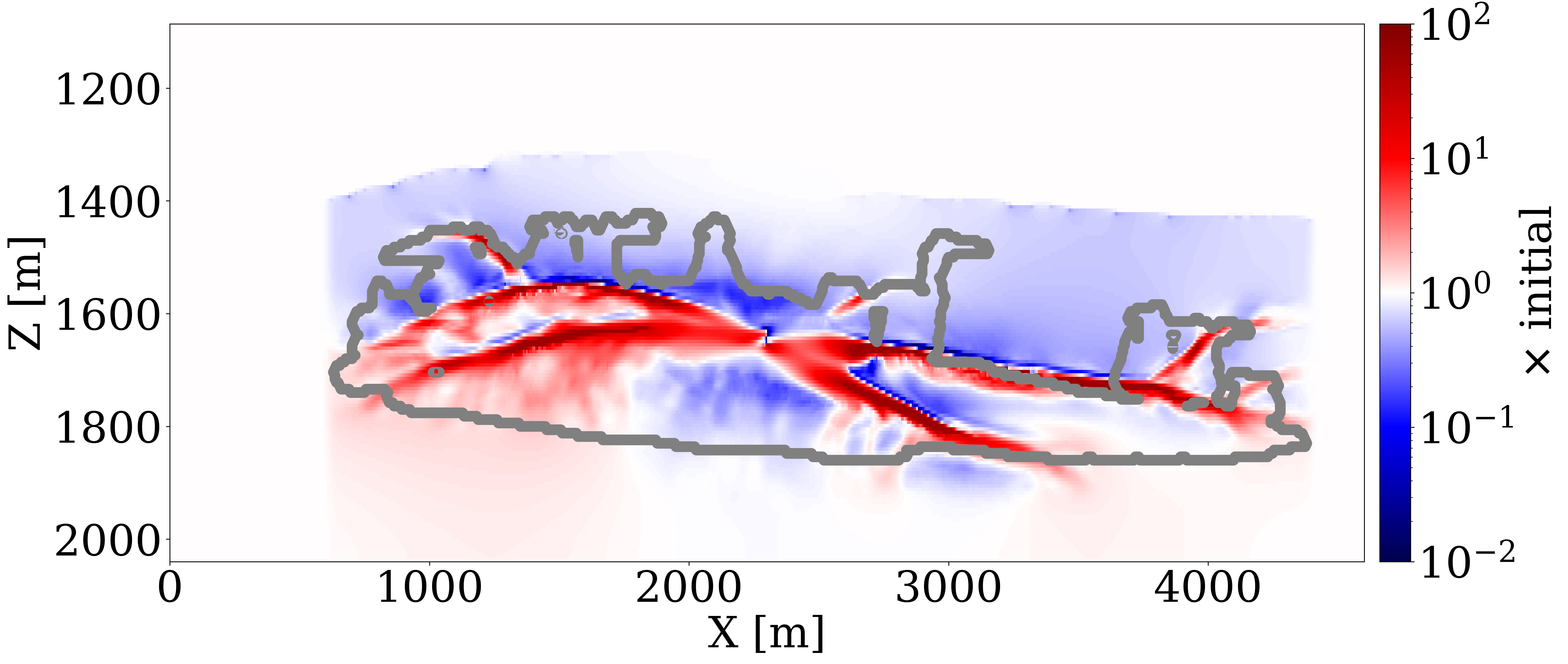}

}

\subcaption{\label{fig-non-inv-crime-perm-update}}

\end{minipage}%

\caption{\label{fig-non-inv-crime}Permeability inversion results for
case 3. (a) Inverted brine-filled baseline velocity used in permeability
inversion. (b) The logarithmic ratio of the inverted permeability to the
initial one. Gray curve indicates the boundary of the
CO\textsubscript{2} plume at the 25th year.}

\end{figure}

To construct an inaccurate but realistic baseline velocity model, we
assume to access to cross-well and reflective seismic data collected at
the baseline stage. By applying 10 datapass of FWI on a smoothed
velocity model, we obtain the inverted velocity model depicted in
Figure~\ref{fig-fwi-v}. This model, albeit imperfect, serves as our
baseline for the brine-filled velocity, which is then utilized in the
end-to-end inversion framework post-CO\textsubscript{2} injection.

For this phase of the study, we employ the initial permeability model
from Figure~\ref{fig-init-perm-1} to assess the impact of modeling
errors on the inversion results. The update to the permeability, shown
in logarithmic scale in Figure~\ref{fig-non-inv-crime-perm-update},
reveals some artifacts outside the CO2 plume area due to the modeling
inaccuracies. Additionally, a high permeability zone within the plume is
slightly misplaced when compared to the updates in
Figure~\ref{fig-case1-update-perm}, yet the overall trend of
permeability changes is correctly captured.

Following this permeability update, we proceed with reservoir
simulations using the updated permeability models to assess the
corrections made to the CO\textsubscript{2} plume predictions. This step
is crucial for validating the practical utility of the inversion
framework for real-world seismic monitoring scenarios.

\subsection{\texorpdfstring{CO\textsubscript{2} plume estimation and
forecast}{CO2 plume estimation and forecast}}\label{co2-plume-estimation-and-forecast}

\begin{figure}

\begin{minipage}{0.33\linewidth}
\includegraphics{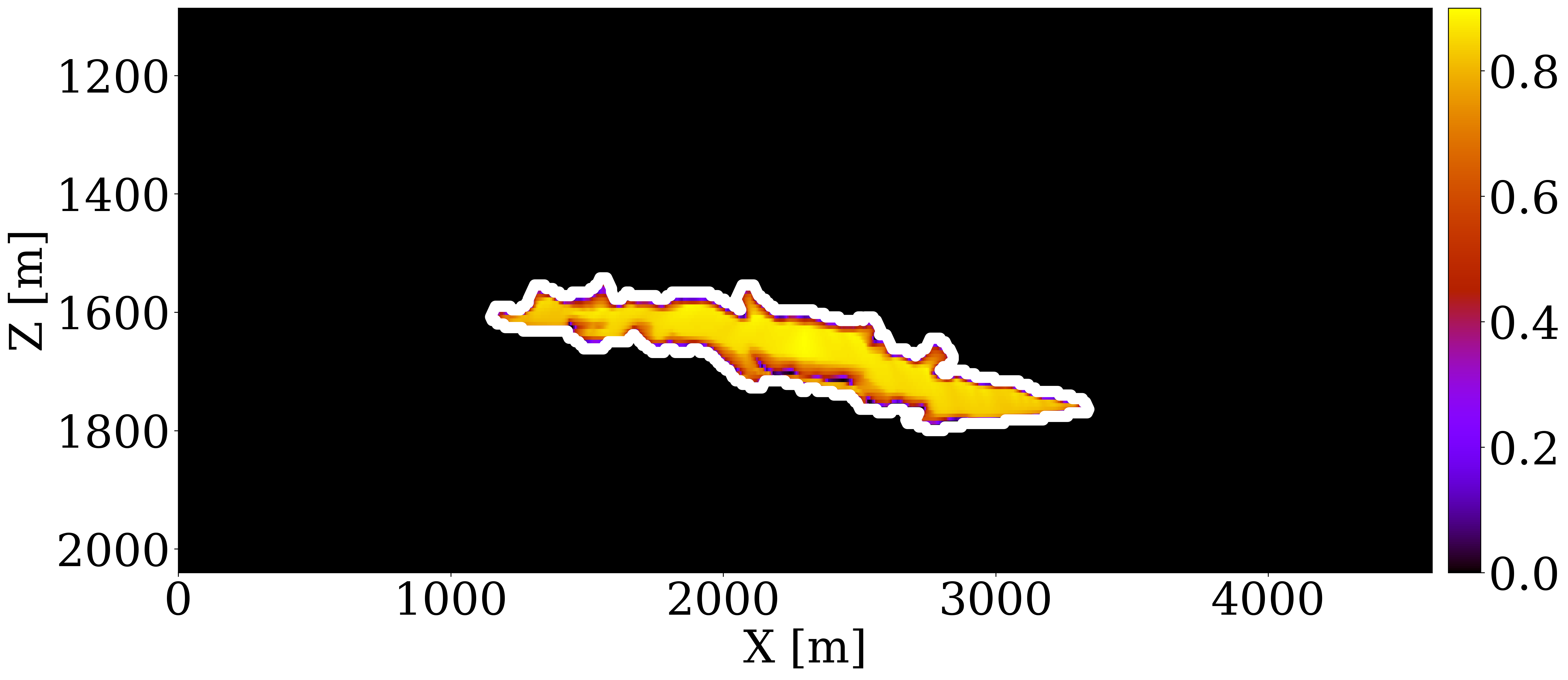}\end{minipage}%
\begin{minipage}{0.33\linewidth}
\includegraphics{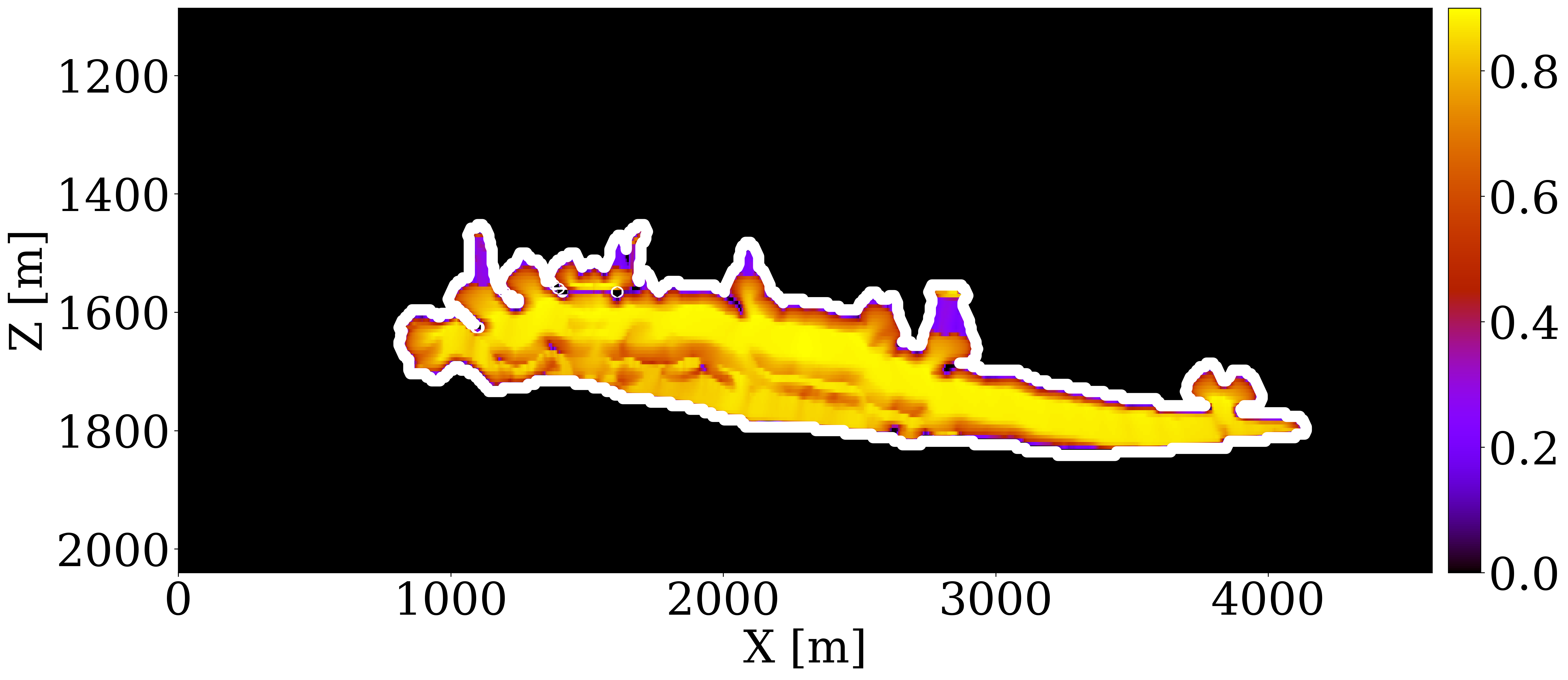}\end{minipage}%
\begin{minipage}{0.33\linewidth}
\includegraphics{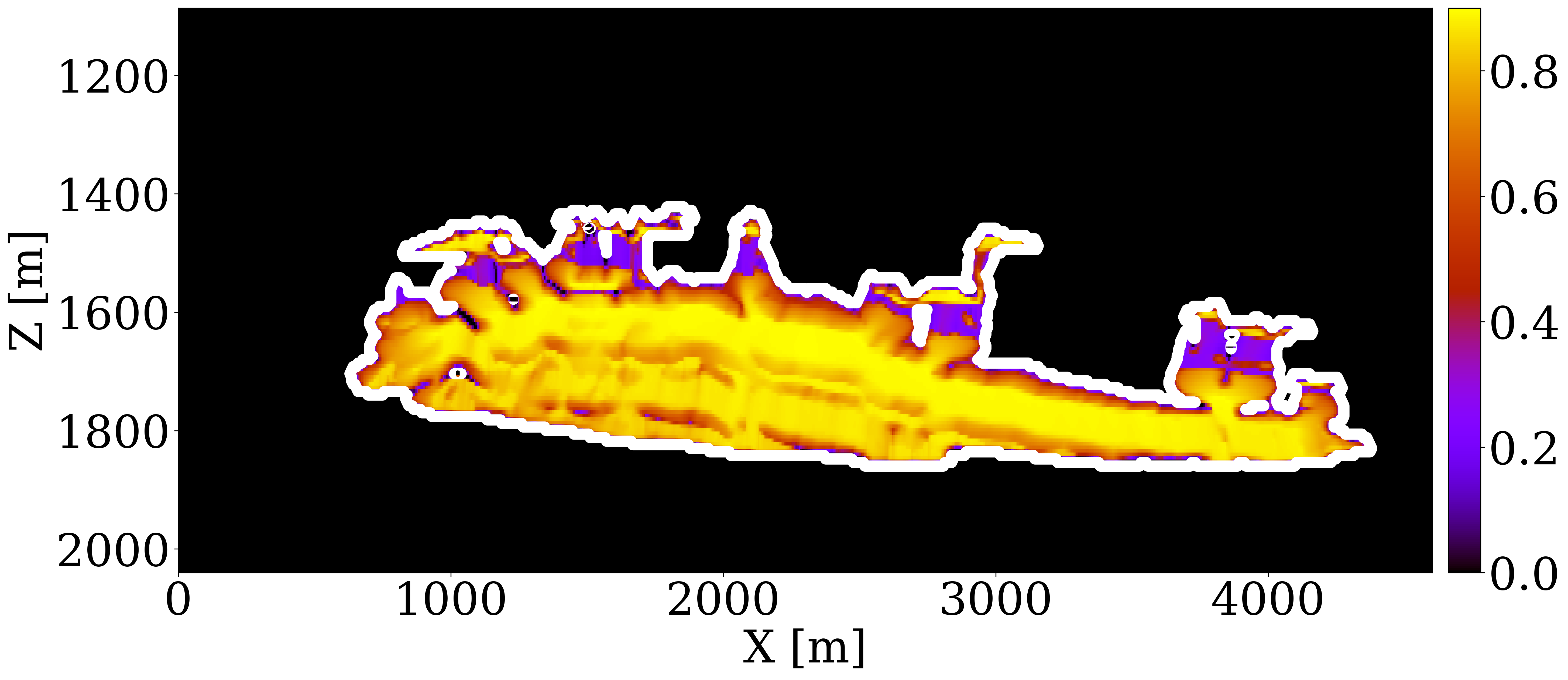}\end{minipage}%
\newline
\begin{minipage}{0.33\linewidth}
\includegraphics{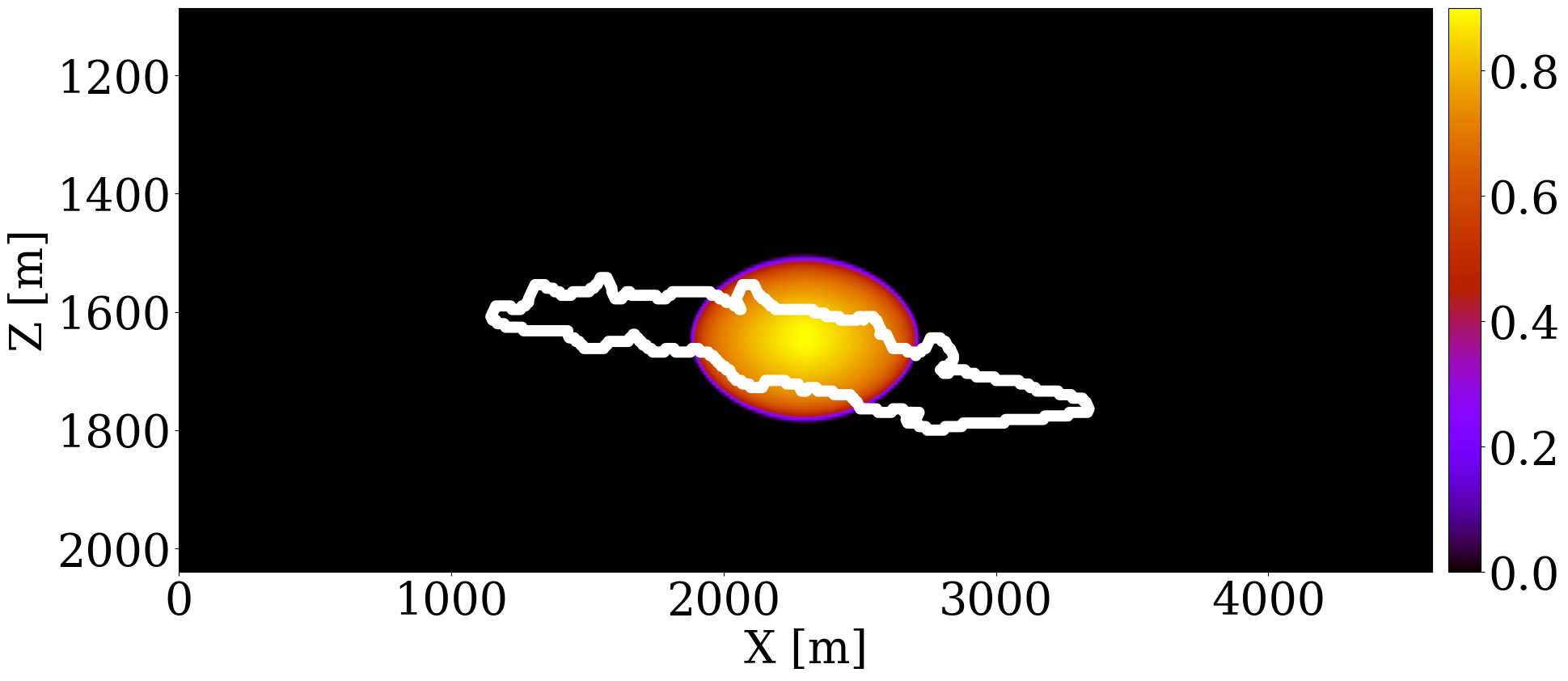}\end{minipage}%
\begin{minipage}{0.33\linewidth}
\includegraphics{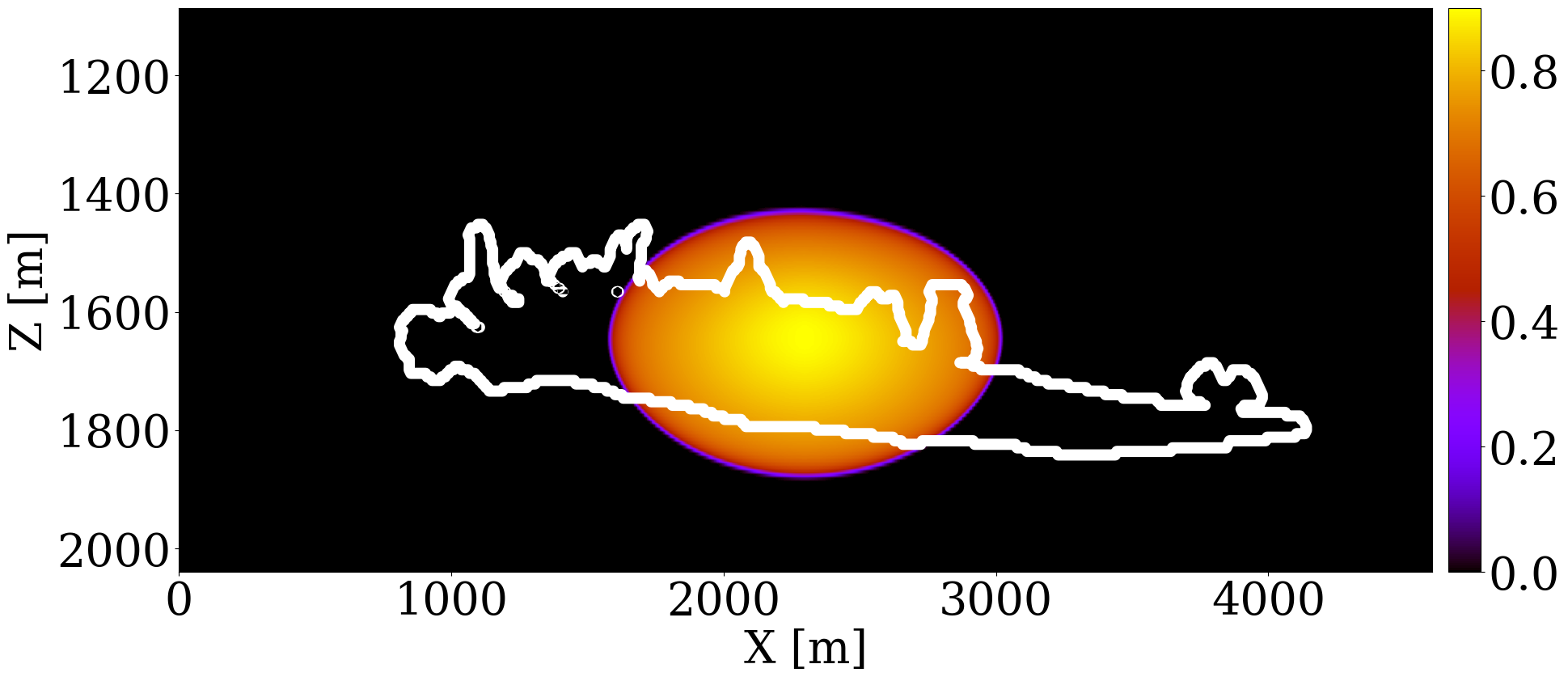}\end{minipage}%
\begin{minipage}{0.33\linewidth}
\includegraphics{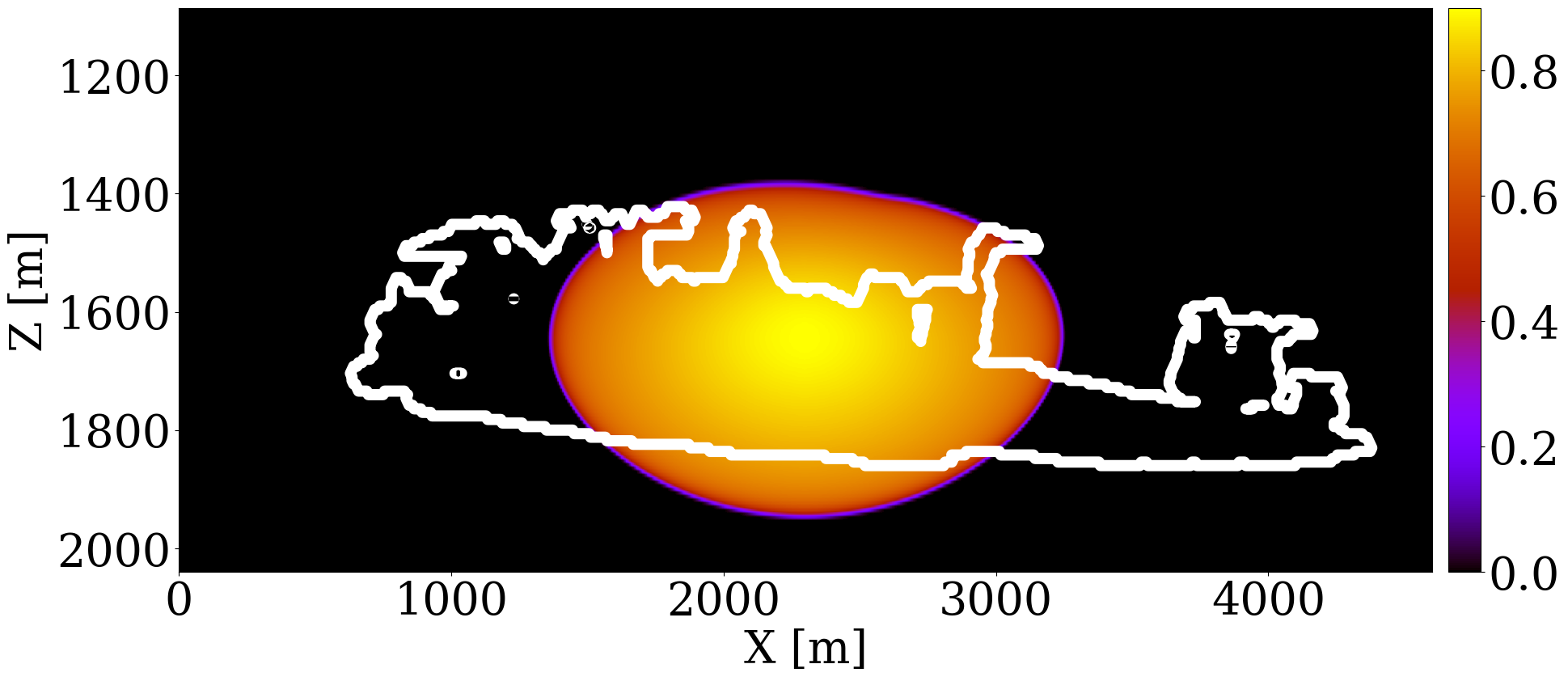}\end{minipage}%
\newline
\begin{minipage}{0.33\linewidth}
\includegraphics{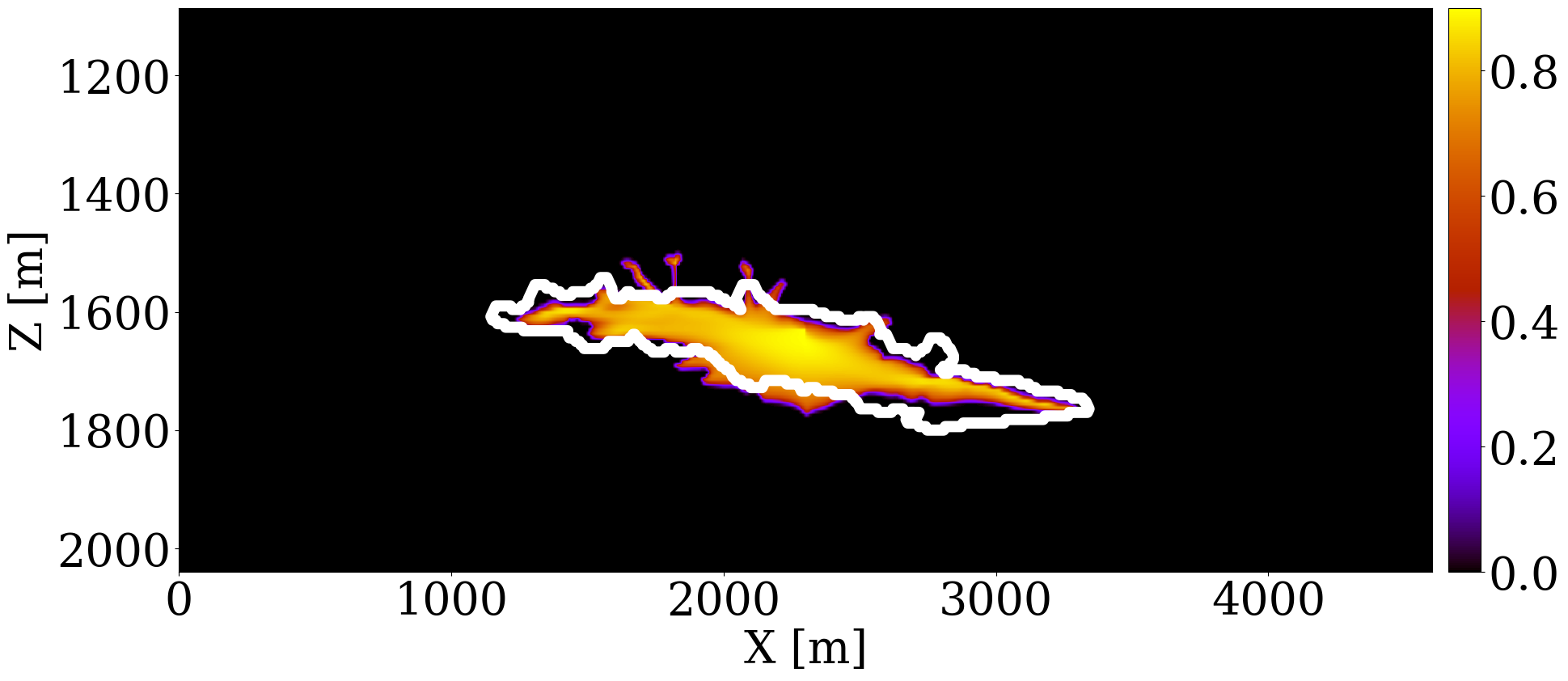}\end{minipage}%
\begin{minipage}{0.33\linewidth}
\includegraphics{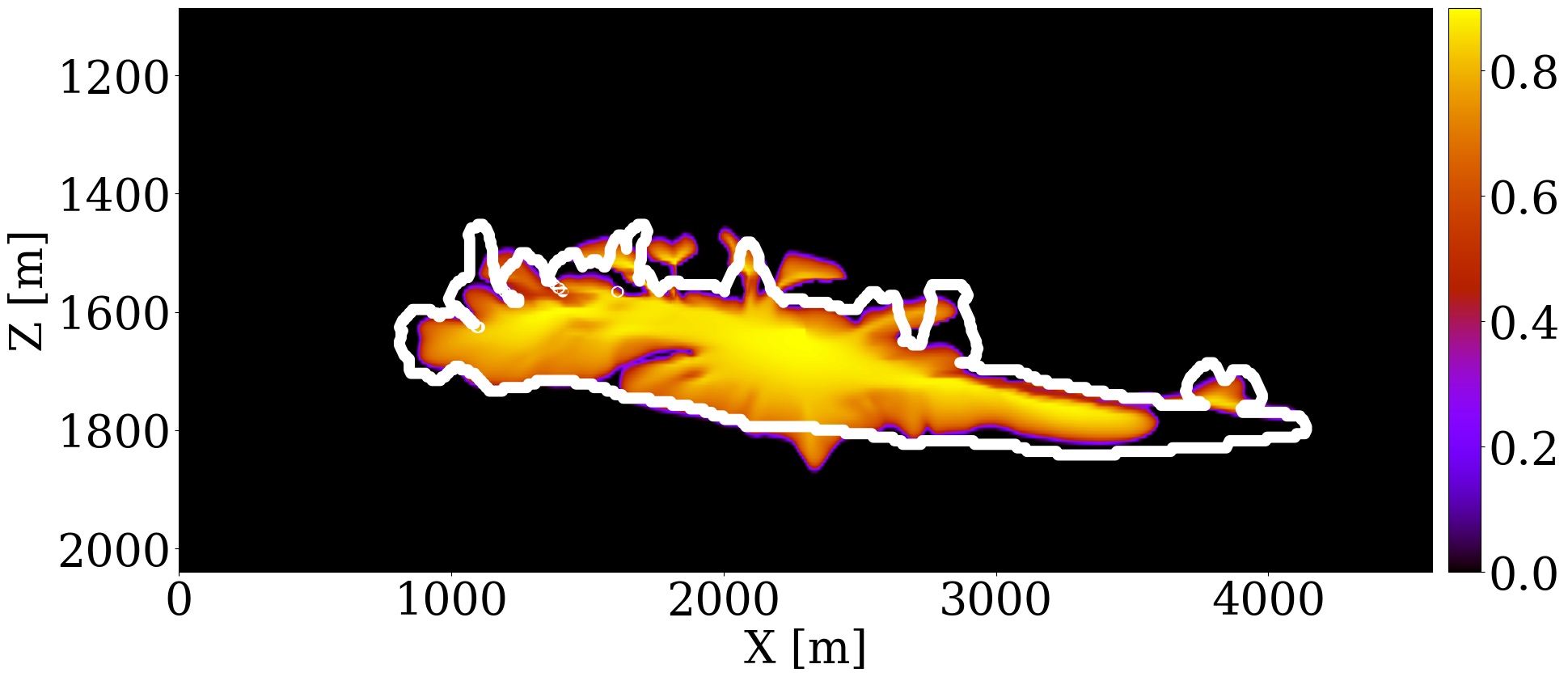}\end{minipage}%
\begin{minipage}{0.33\linewidth}
\includegraphics{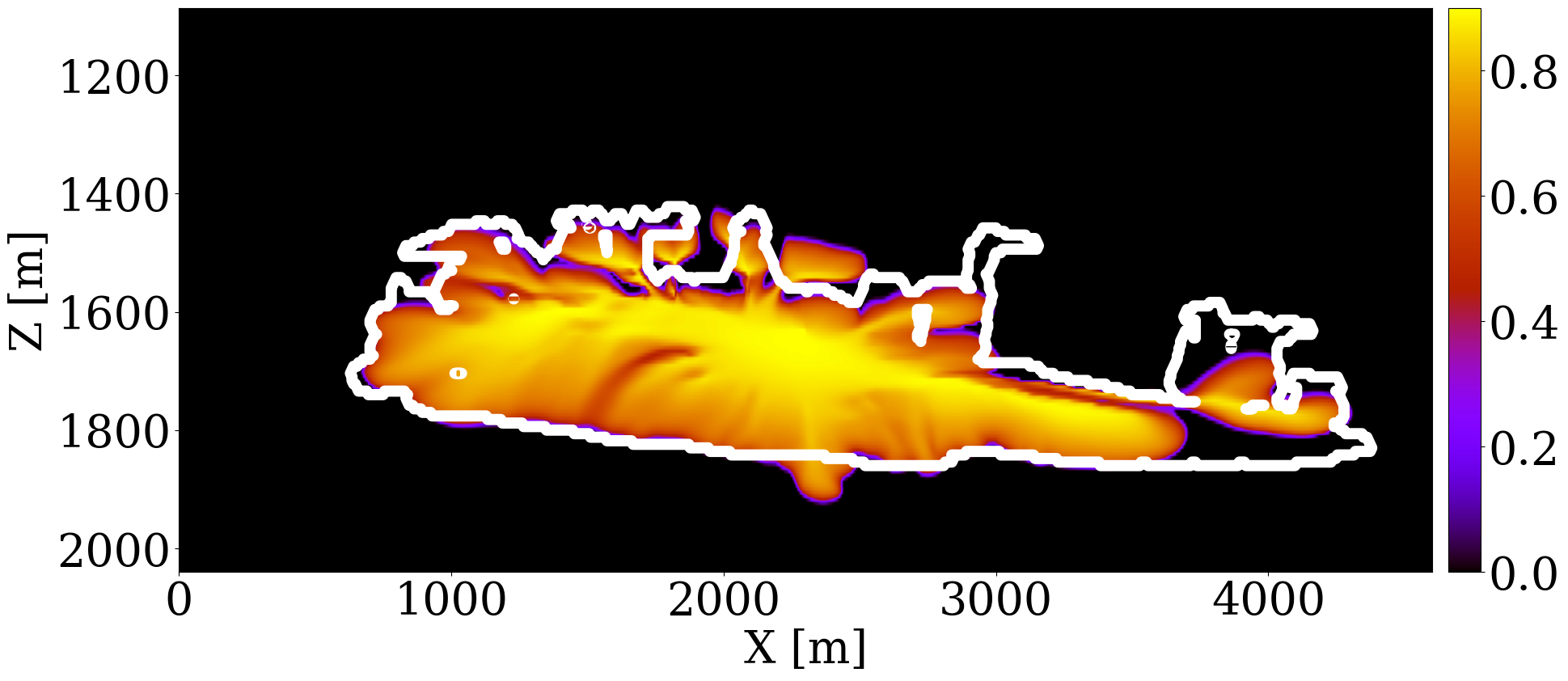}\end{minipage}%
\newline
\begin{minipage}{0.33\linewidth}
\includegraphics{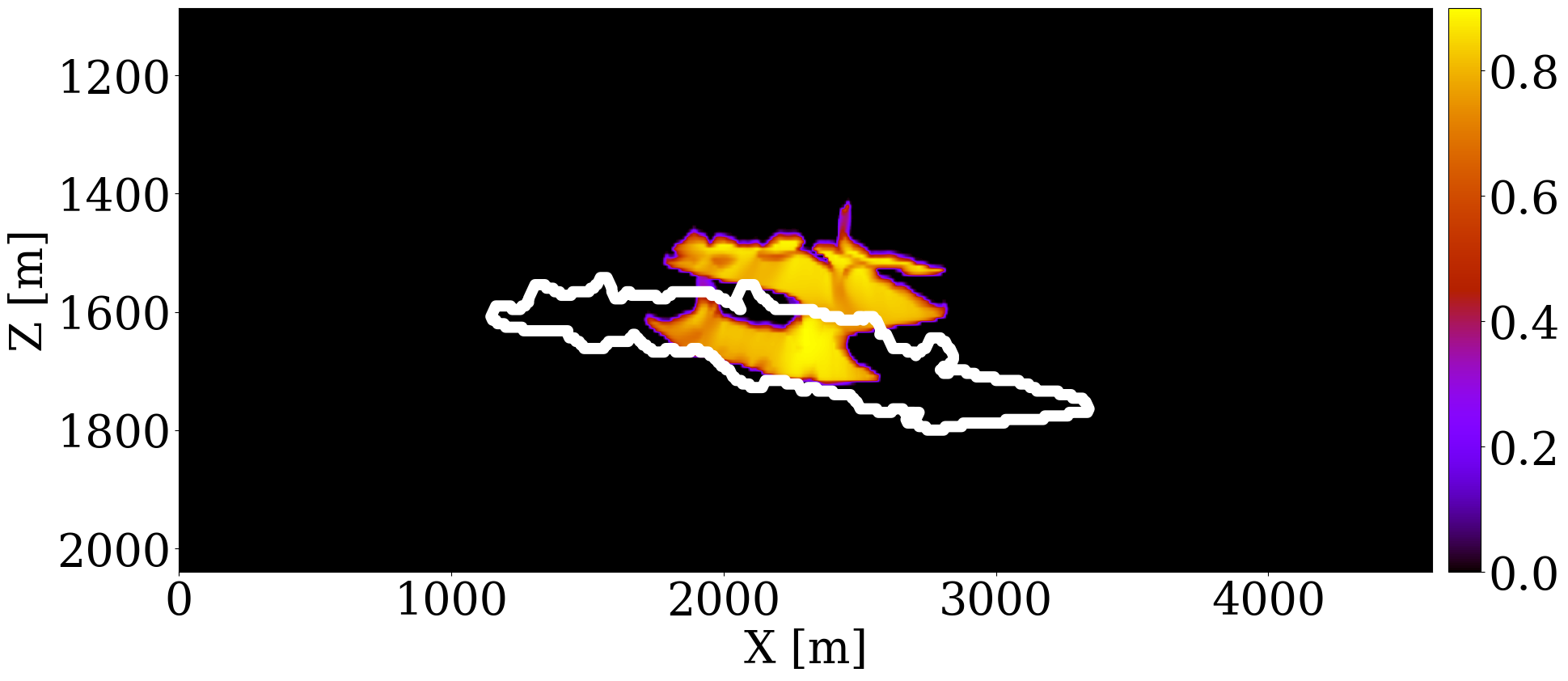}\end{minipage}%
\begin{minipage}{0.33\linewidth}
\includegraphics{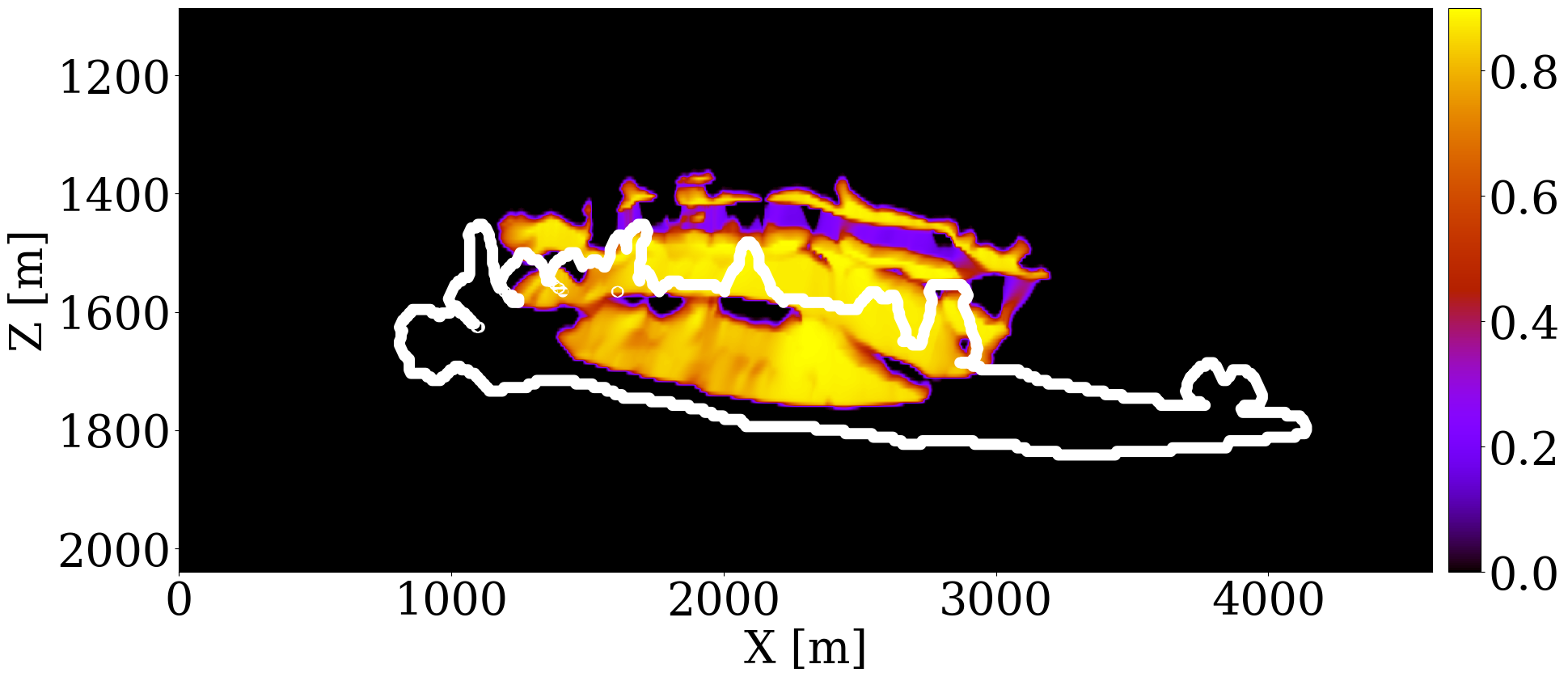}\end{minipage}%
\begin{minipage}{0.33\linewidth}
\includegraphics{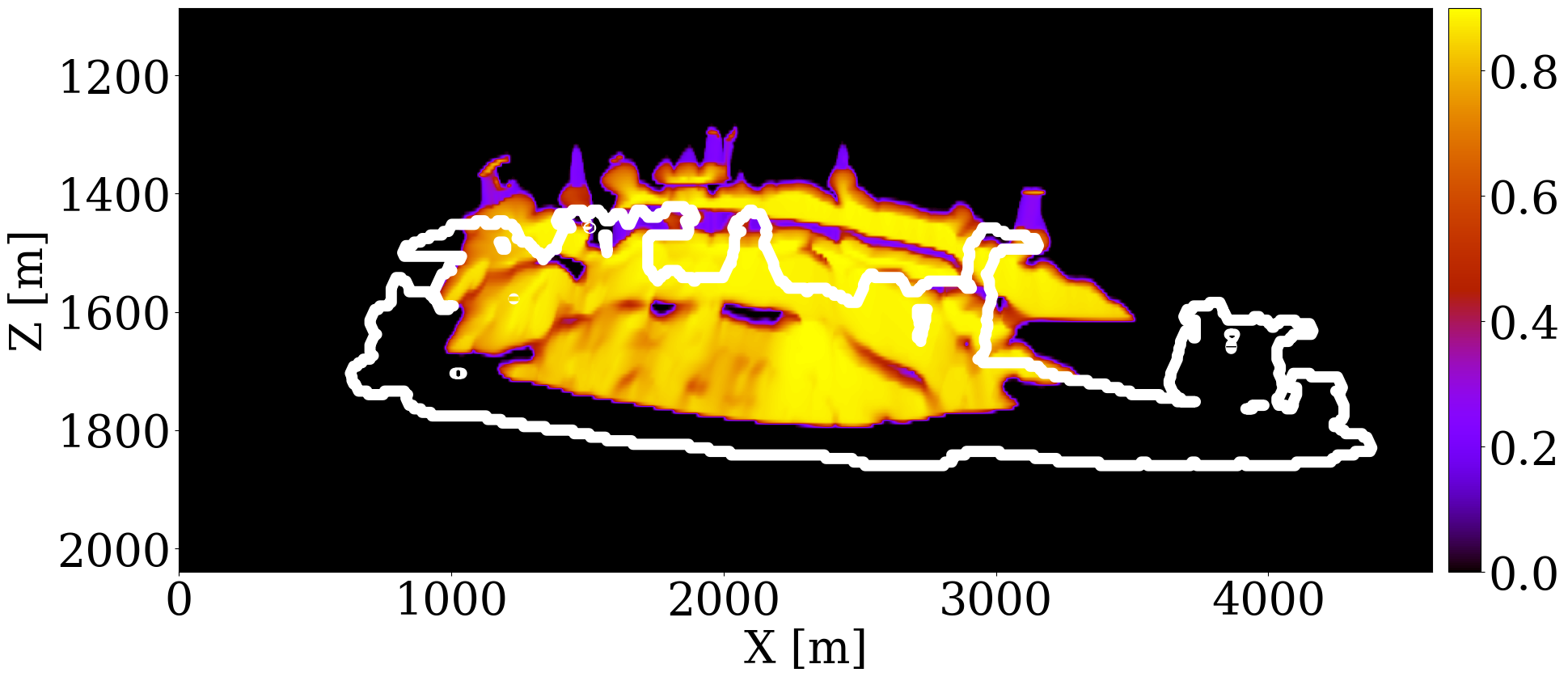}\end{minipage}%
\newline
\begin{minipage}{0.33\linewidth}
\includegraphics{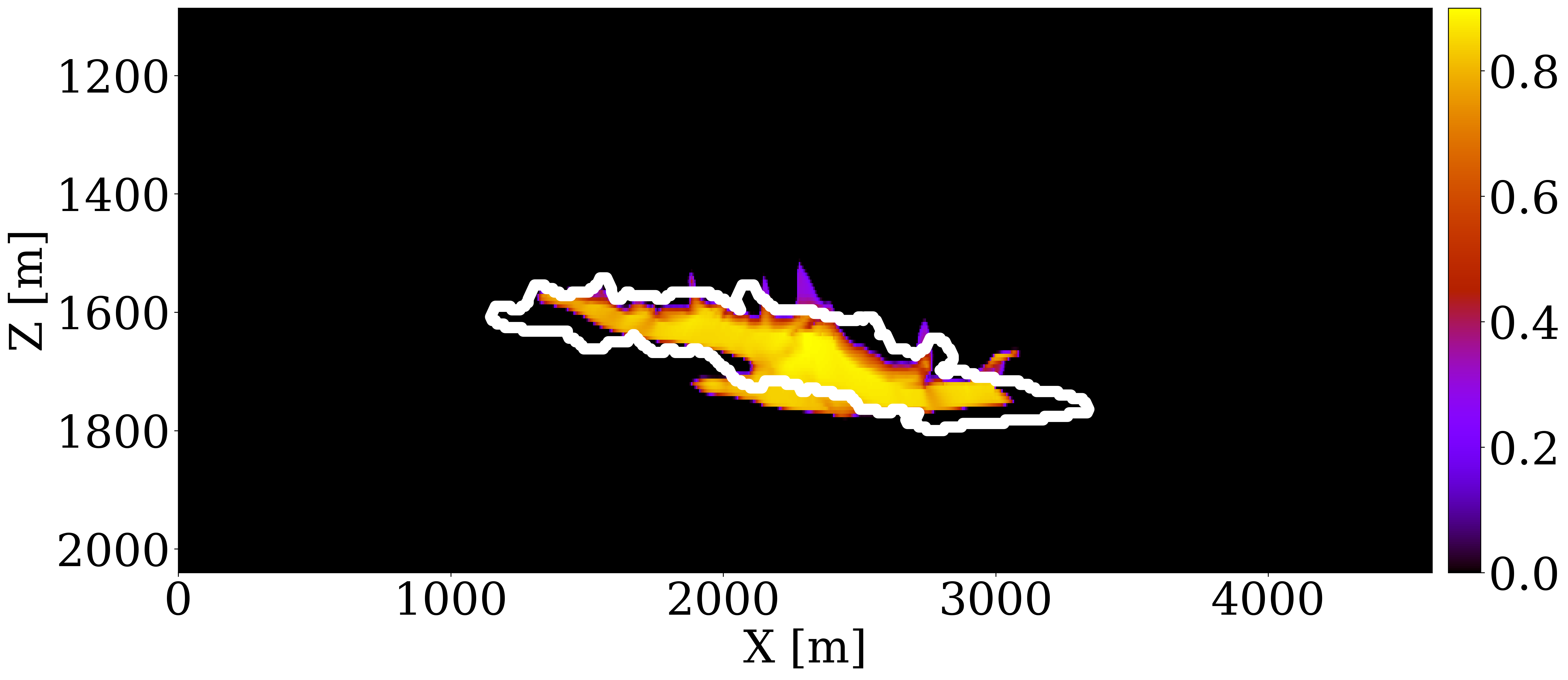}\end{minipage}%
\begin{minipage}{0.33\linewidth}
\includegraphics{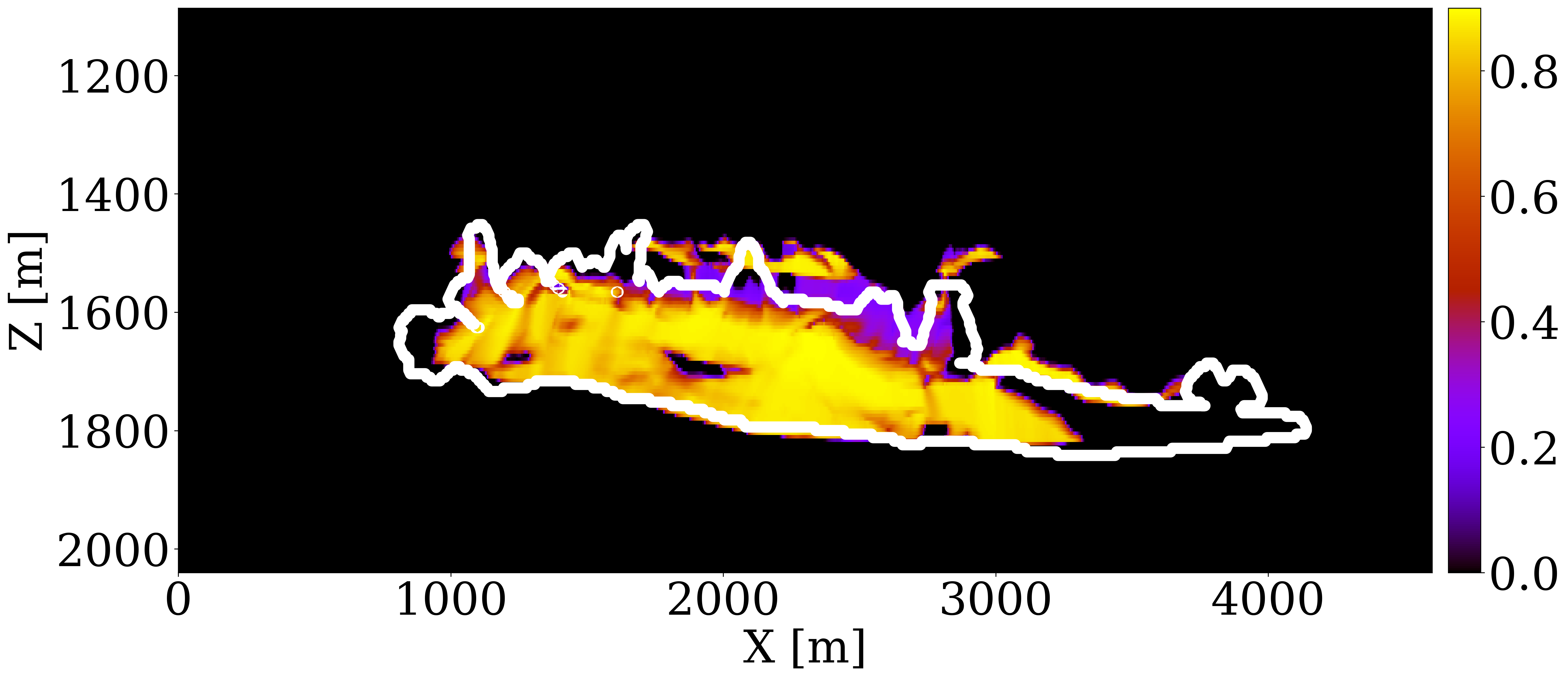}\end{minipage}%
\begin{minipage}{0.33\linewidth}
\includegraphics{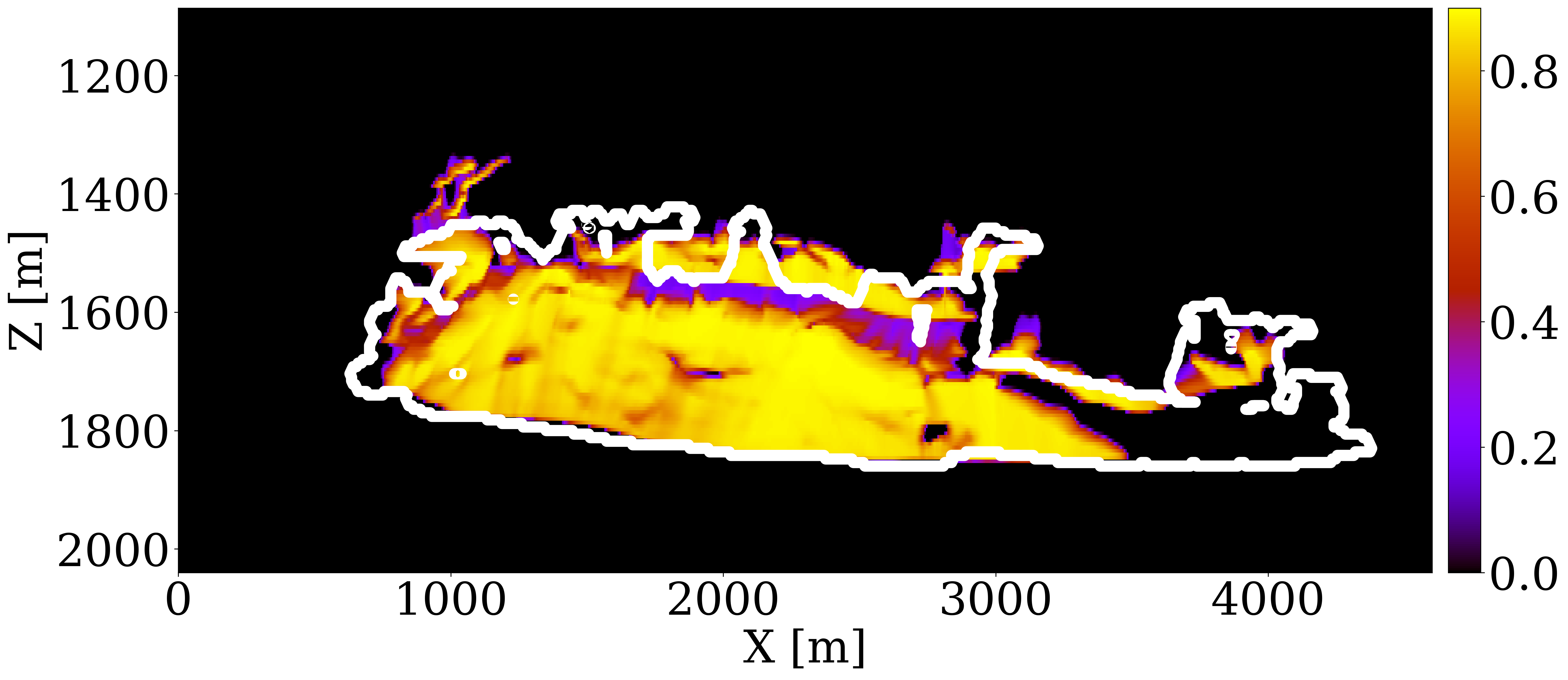}\end{minipage}%
\newline
\begin{minipage}{0.33\linewidth}
\includegraphics{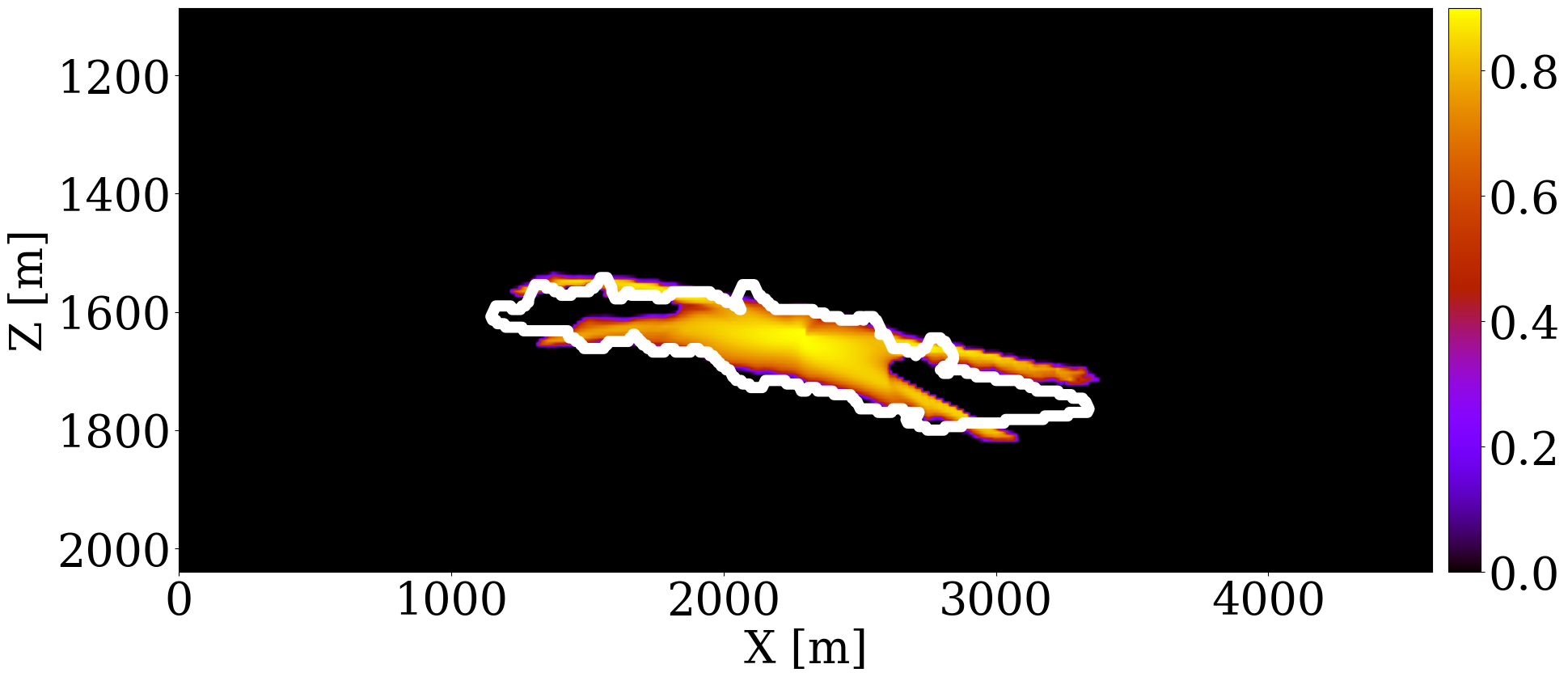}\end{minipage}%
\begin{minipage}{0.33\linewidth}
\includegraphics{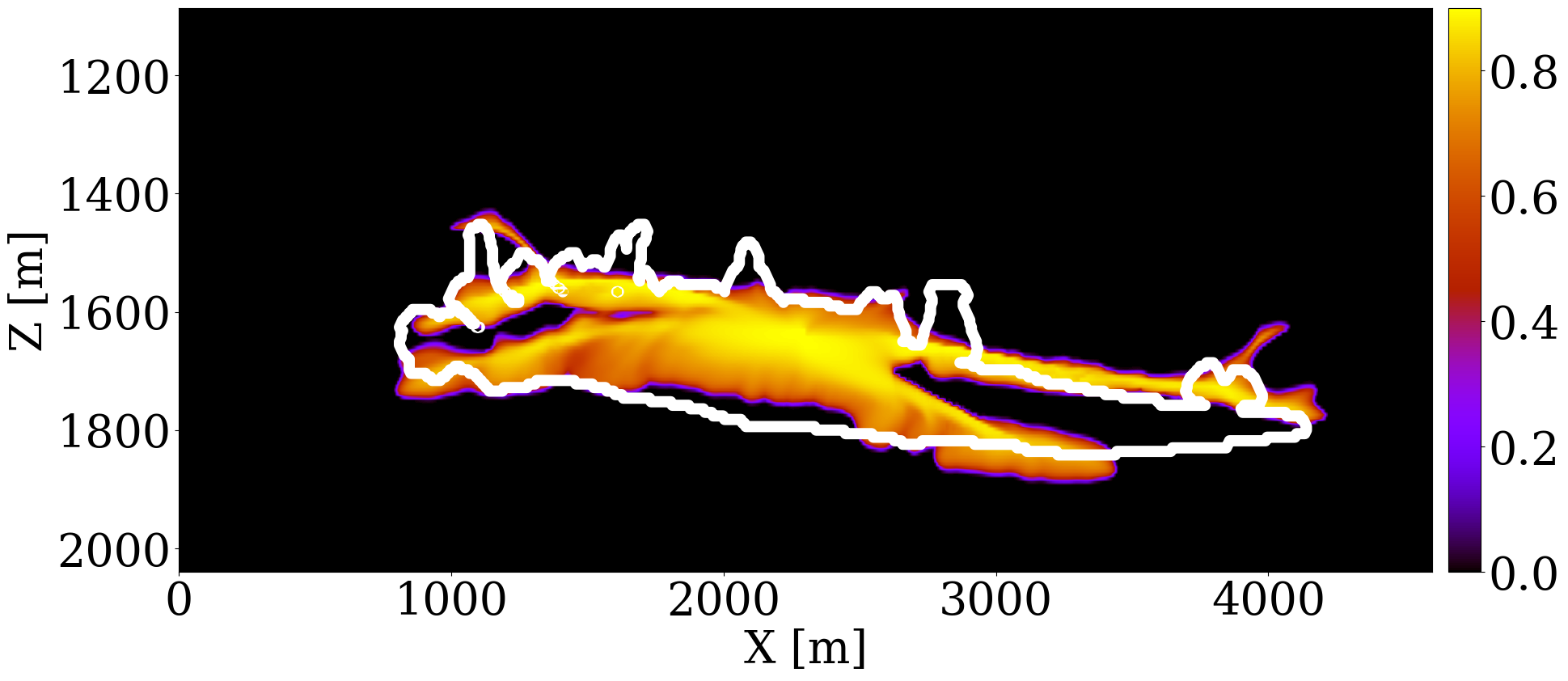}\end{minipage}%
\begin{minipage}{0.33\linewidth}
\includegraphics{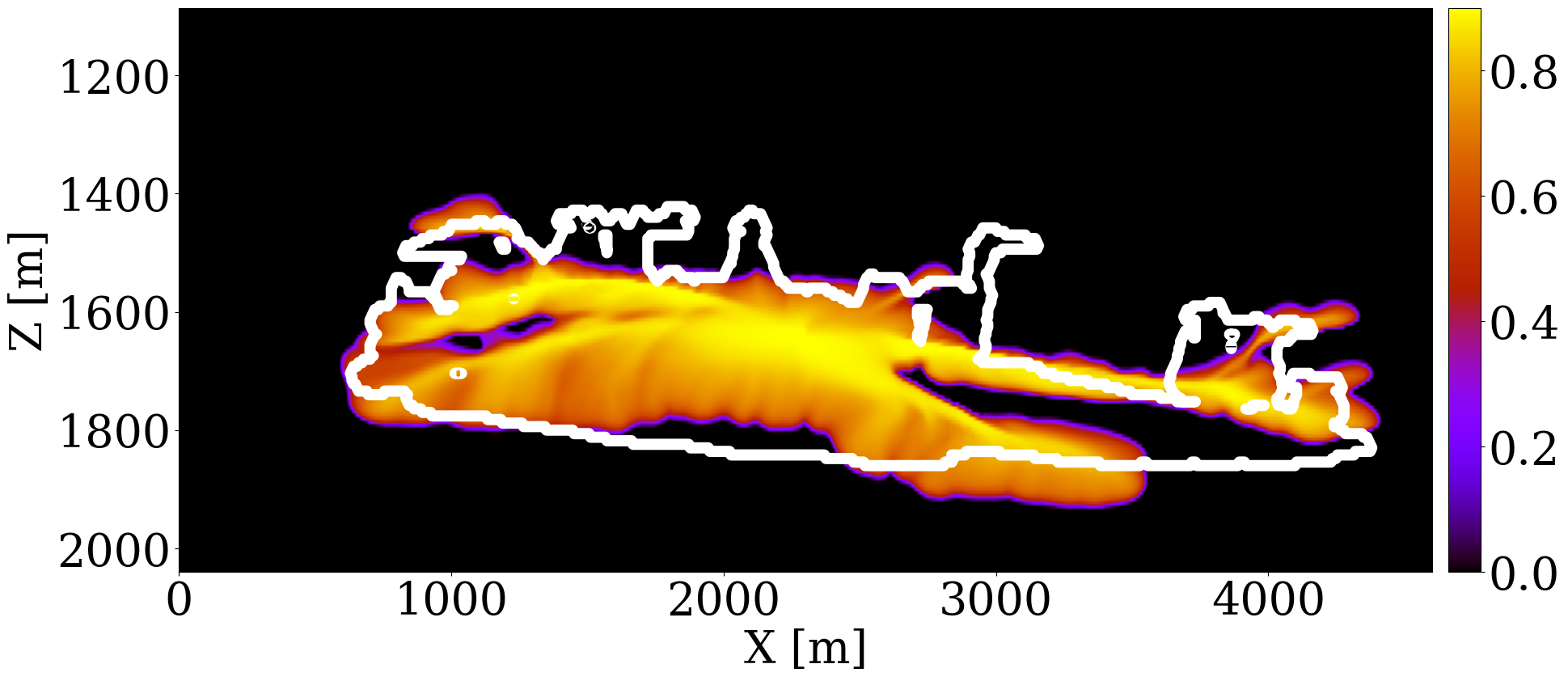}\end{minipage}%

\caption{\label{fig-co2}Predicted CO\textsubscript{2} saturation for
5th, 15th, and 25th years, shown in first, second, and third columns,
respectively. The first row shows the (unseen) ground truth
CO\textsubscript{2} saturation. The second and fourth rows show the
saturation predicted with initial permeability models in
Figure~\ref{fig-init-perm-1} and Figure~\ref{fig-init-perm-2},
respectively. The third, fifth, and sixth rows show the updated
saturation after updating the initial permeability models by
Figure~\ref{fig-case1-update-perm}, Figure~\ref{fig-case2-update-perm},
and Figure~\ref{fig-non-inv-crime-perm-update}, respectively. The
boundaries of the (unseen) ground truth CO\textsubscript{2} saturation
are shown in white curves.}

\end{figure}

The primary objective of our end-to-end inversion framework is to
accurately estimate reservoir permeability, a crucial step towards the
ultimate goal of predicting CO\textsubscript{2} saturation both
historically and in the near future for effective GCS monitoring. To
validate the framework's efficacy, we conduct a quality control
comparison involving CO\textsubscript{2} saturation simulations based on
initial, inverted, and ground truth permeability models, as depicted in
Figure~\ref{fig-co2}. Across all simulations, we note substantial
improvements in the CO\textsubscript{2} plume shape predictions, closely
aligning with the boundaries of the ground truth CO\textsubscript{2}
plume. Notably, the initial simulations significantly misjudged the
lateral spread of the CO\textsubscript{2} plume. The corrections applied
through the updated permeability models, however, yield accurate
representations of the plume's lateral extent.

\begin{figure}

\begin{minipage}{0.50\linewidth}
\includegraphics{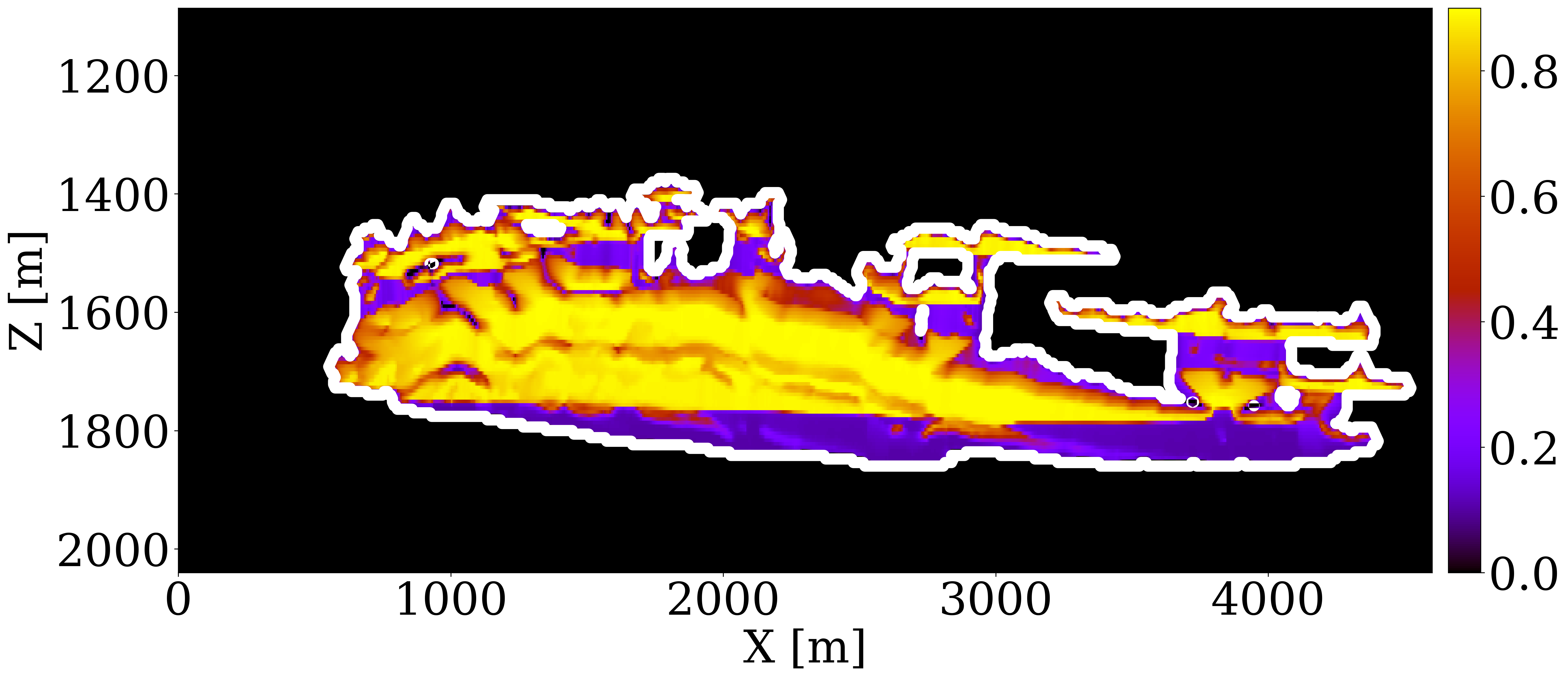}\end{minipage}%
\begin{minipage}{0.50\linewidth}
\includegraphics{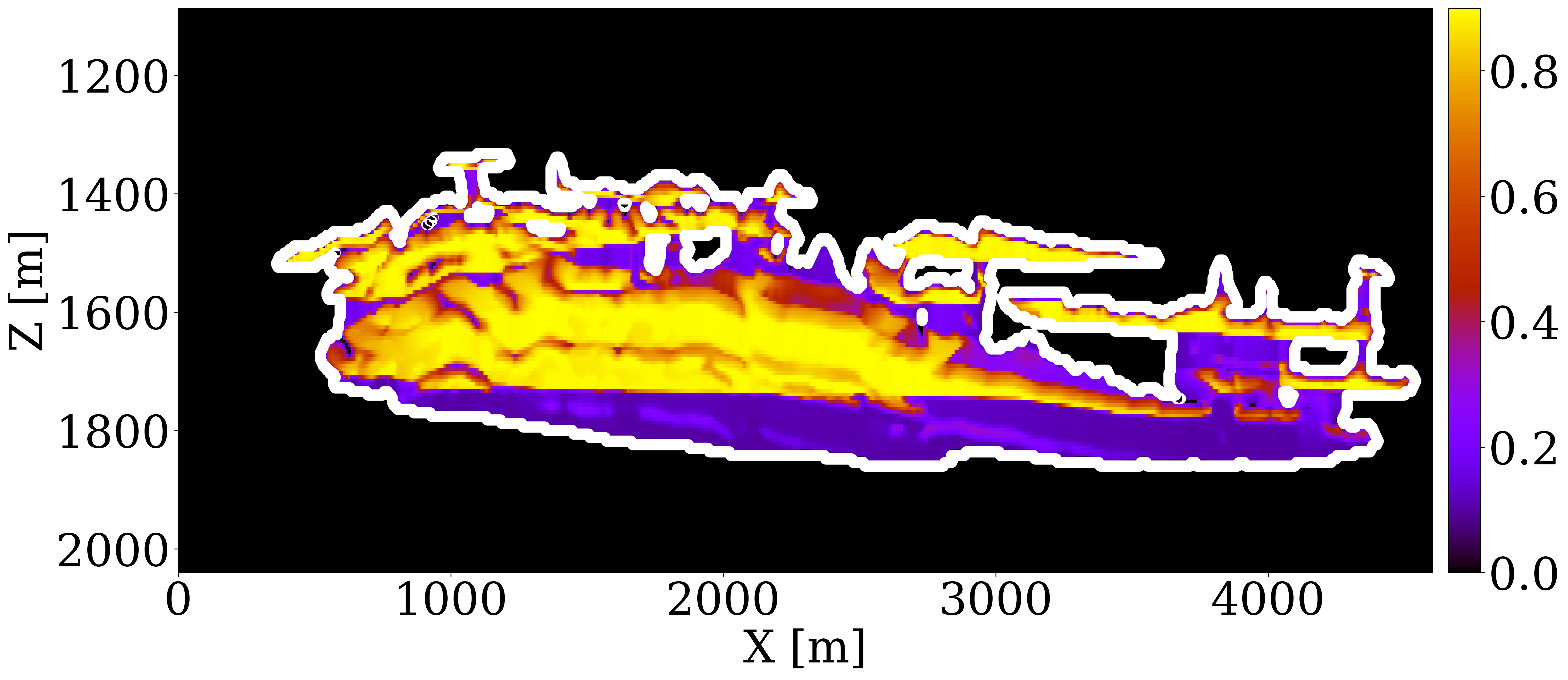}\end{minipage}%
\newline
\begin{minipage}{0.50\linewidth}
\includegraphics{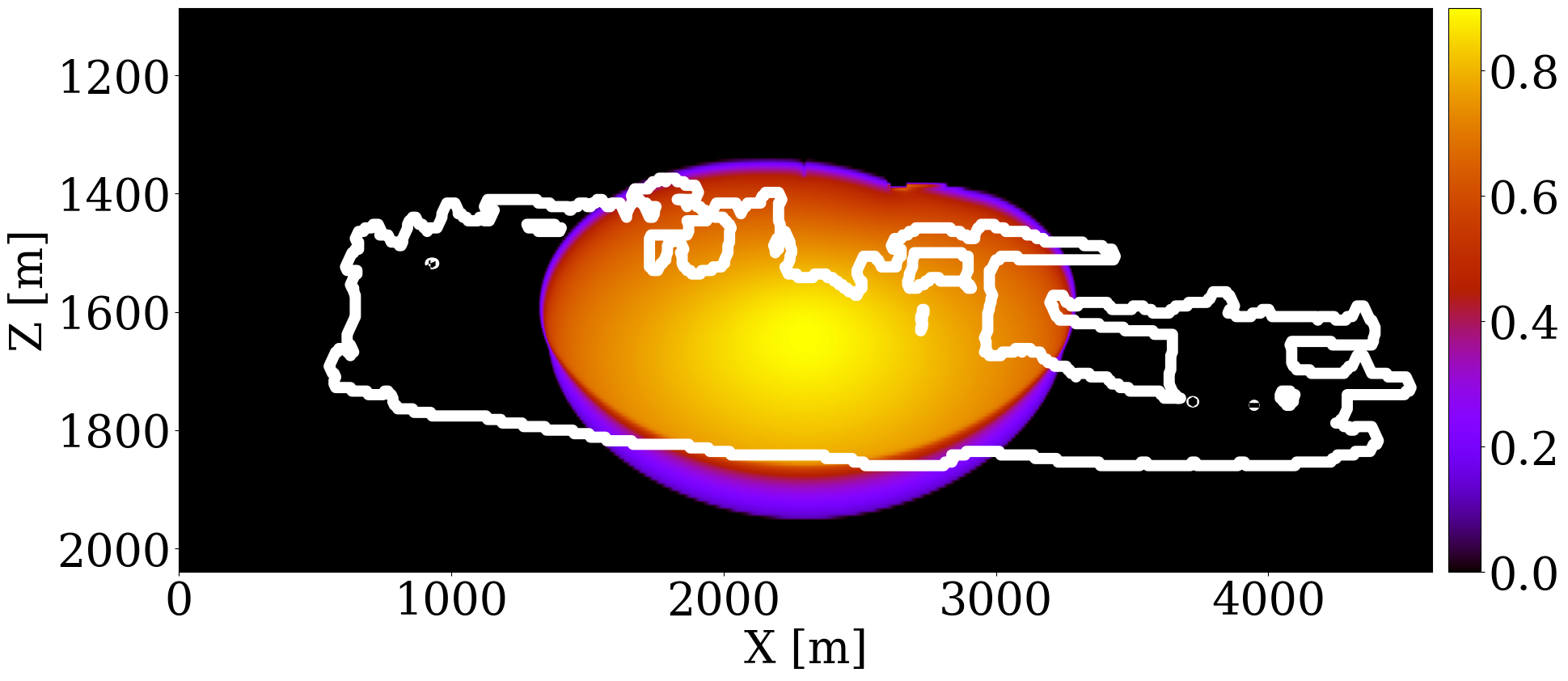}\end{minipage}%
\begin{minipage}{0.50\linewidth}
\includegraphics{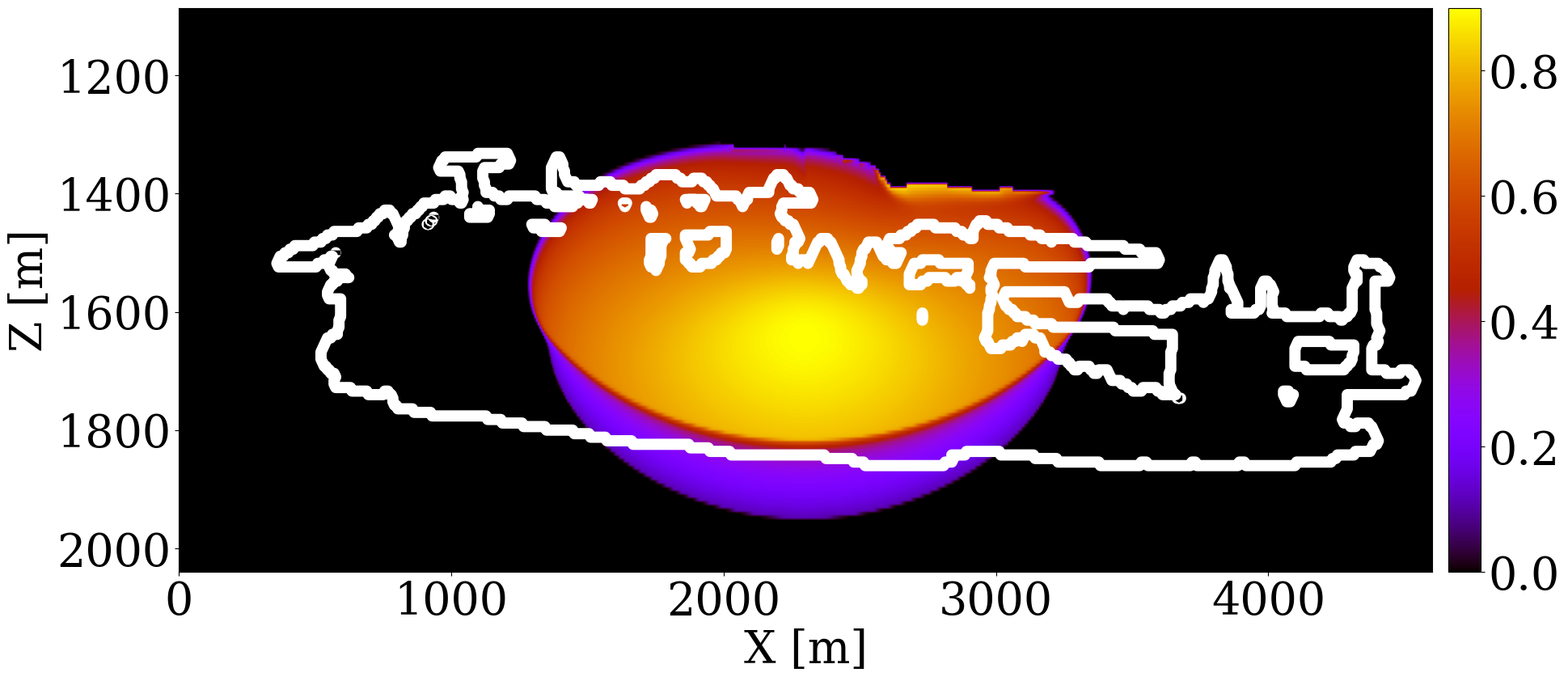}\end{minipage}%
\newline
\begin{minipage}{0.50\linewidth}
\includegraphics{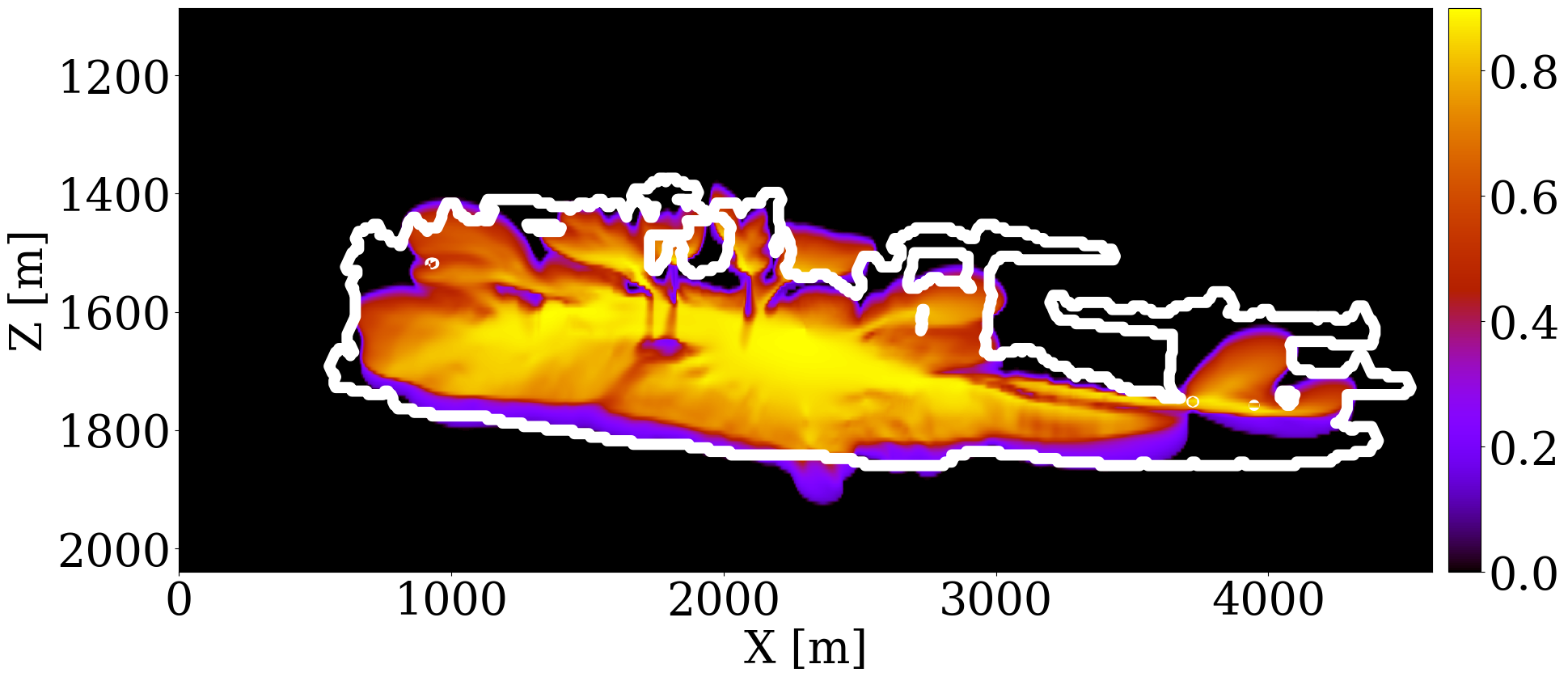}\end{minipage}%
\begin{minipage}{0.50\linewidth}
\includegraphics{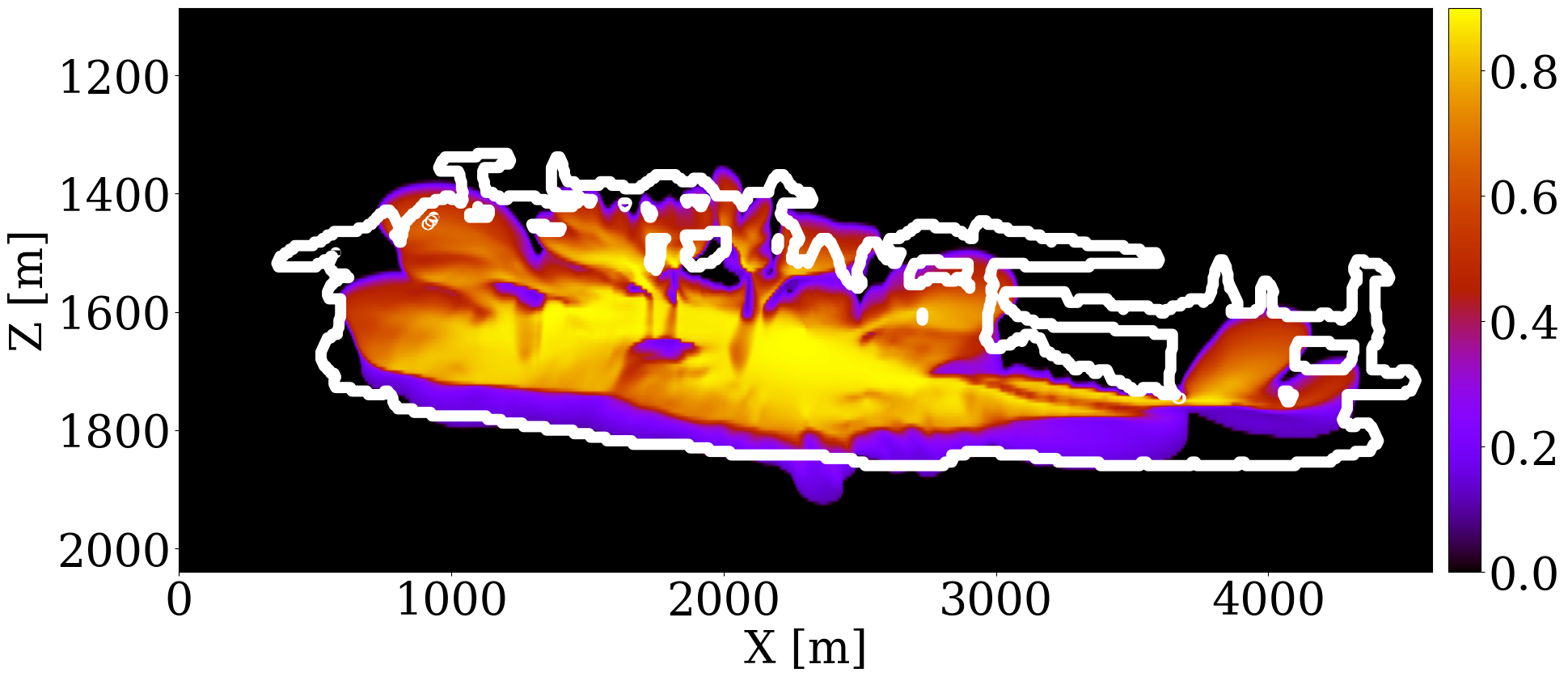}\end{minipage}%
\newline
\begin{minipage}{0.50\linewidth}
\includegraphics{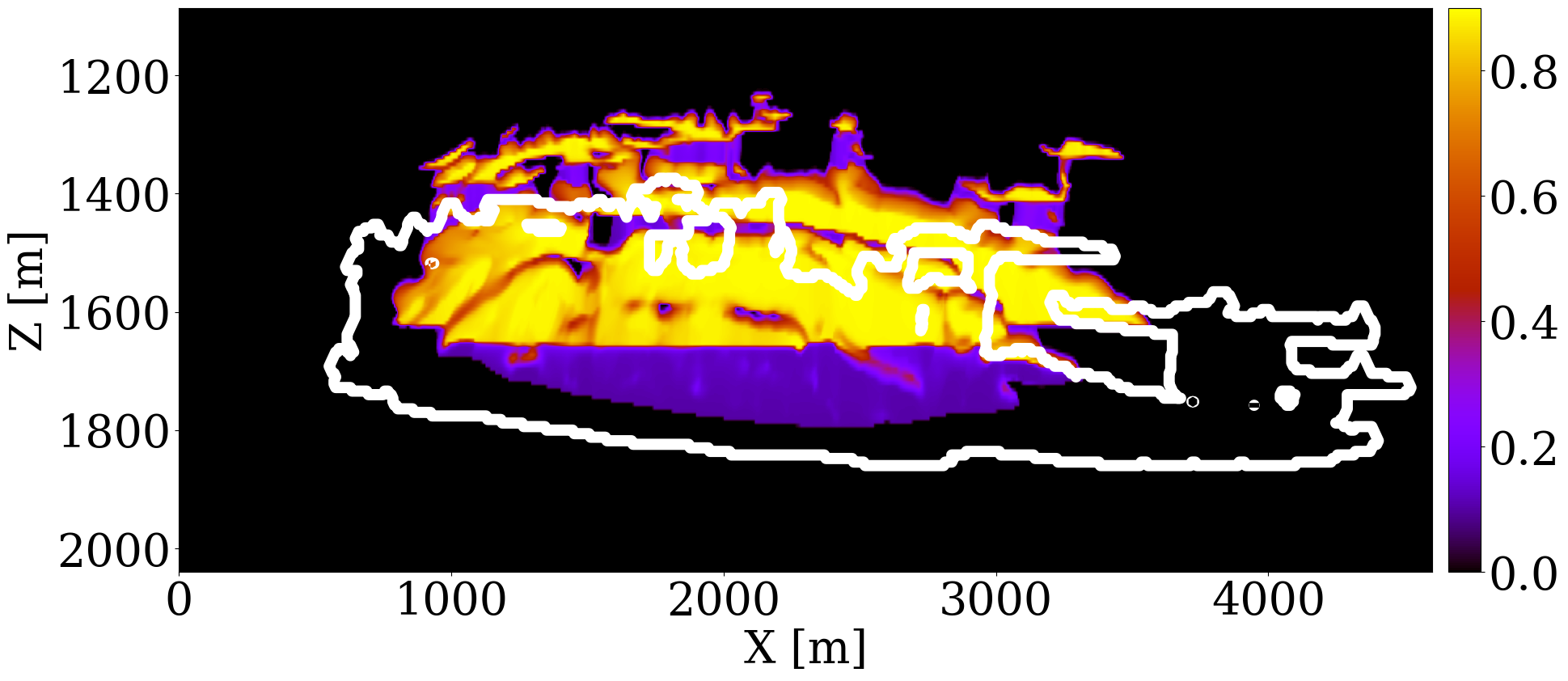}\end{minipage}%
\begin{minipage}{0.50\linewidth}
\includegraphics{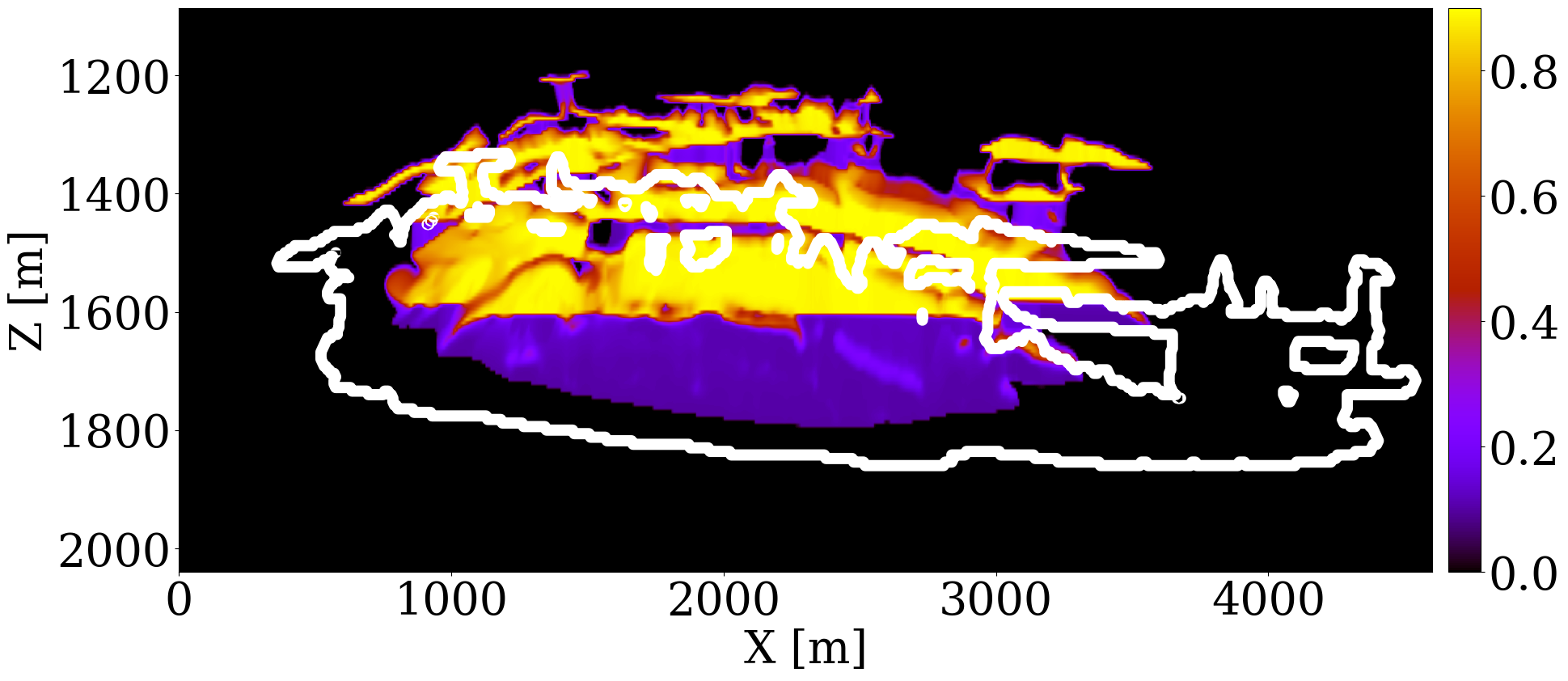}\end{minipage}%
\newline
\begin{minipage}{0.50\linewidth}
\includegraphics{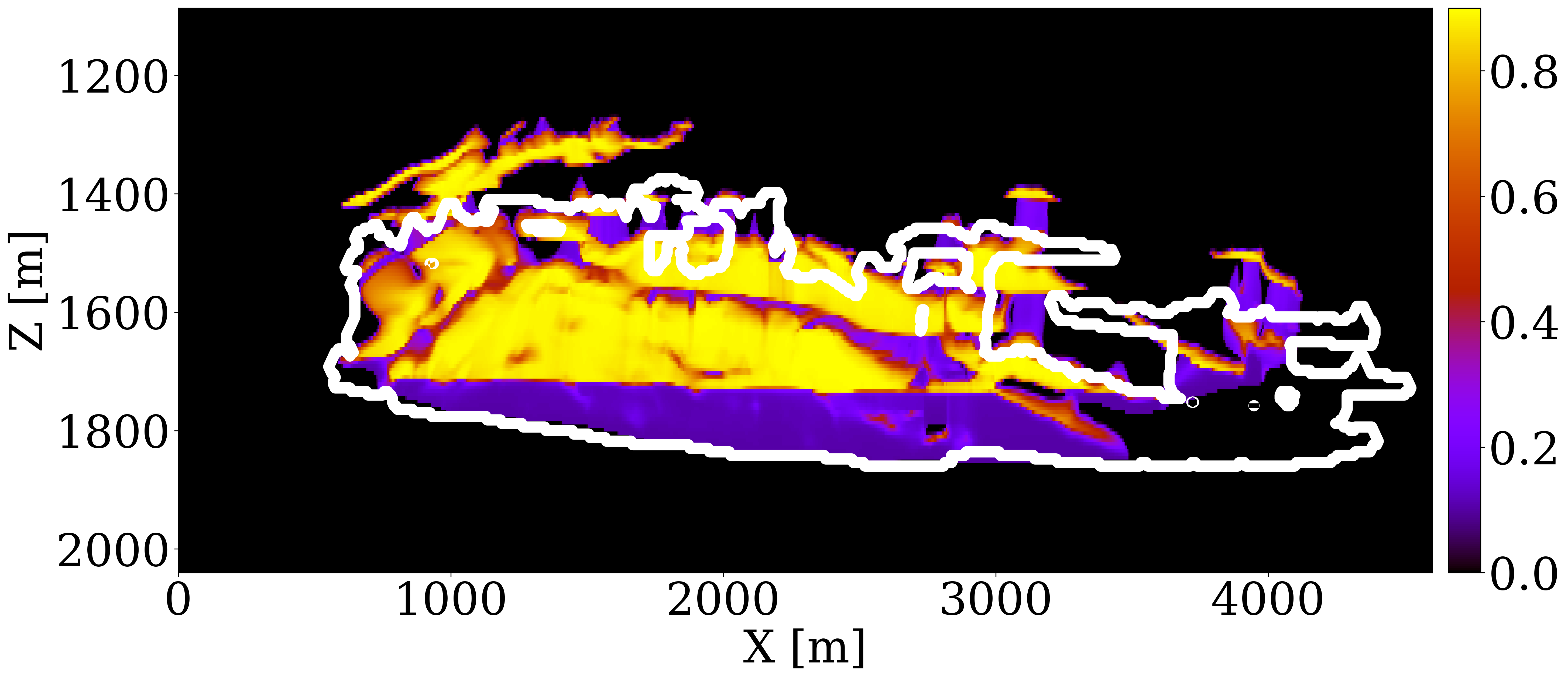}\end{minipage}%
\begin{minipage}{0.50\linewidth}
\includegraphics{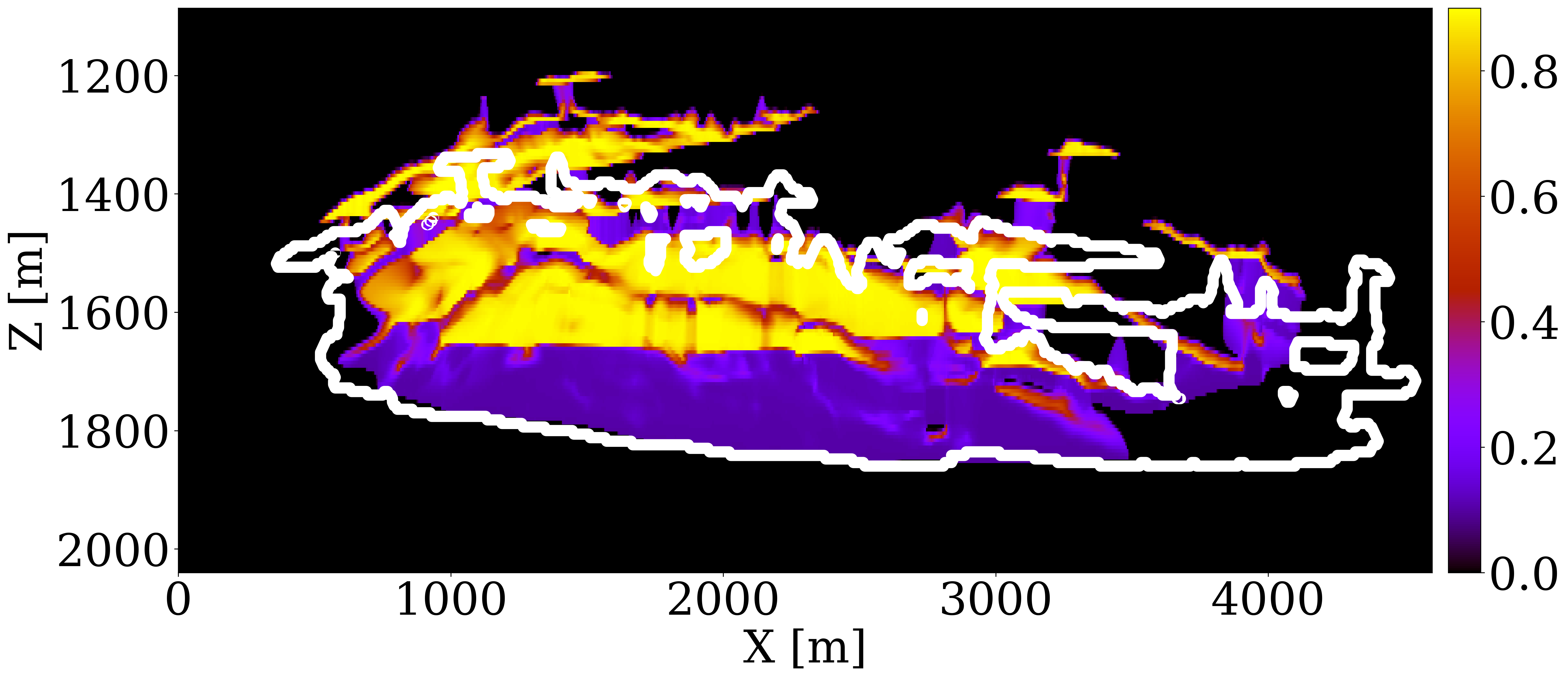}\end{minipage}%
\newline
\begin{minipage}{0.50\linewidth}
\includegraphics{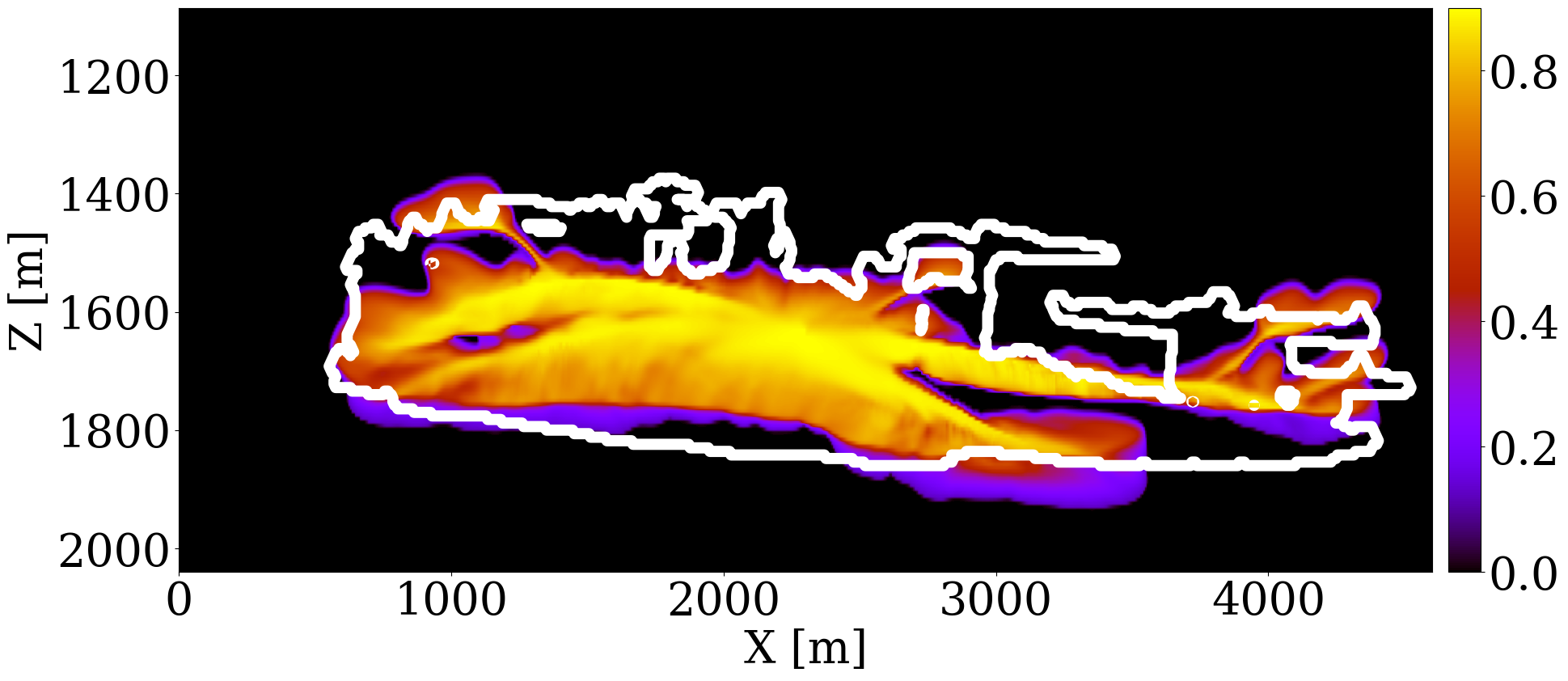}\end{minipage}%
\begin{minipage}{0.50\linewidth}
\includegraphics{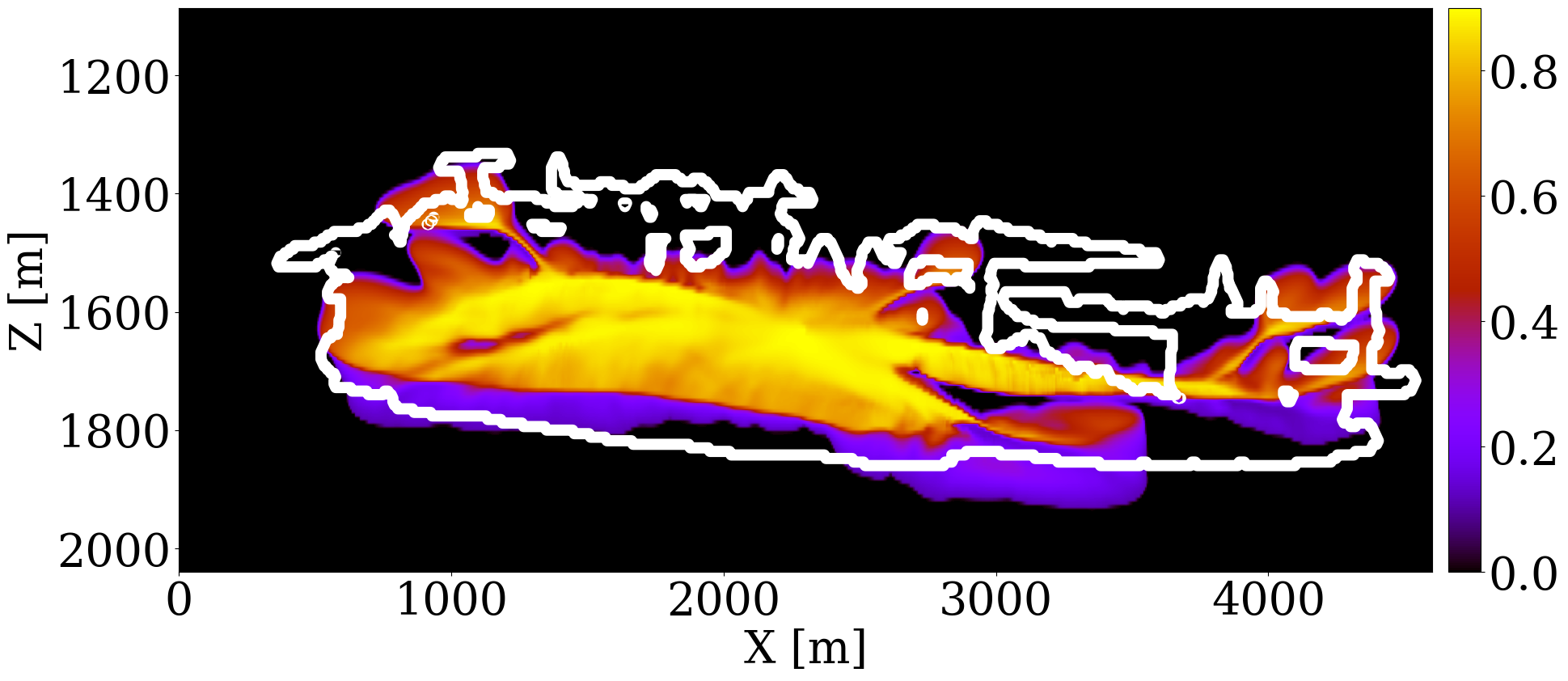}\end{minipage}%

\caption{\label{fig-co2-forecast}CO\textsubscript{2} plume forecasts for
45th and 65th years, shown in first and second columns, respectively.
The ordering of the rows remains the same as Figure~\ref{fig-co2}.
Purple regions display the CO\textsubscript{2} plume permanently stored
via the residual trapping mechanism.}

\end{figure}

Expanding our analysis to future forecasting,
Figure~\ref{fig-co2-forecast} illustrates the predicted movement of the
CO\textsubscript{2} plume over a 40-year period following a 25-year
injection phase, without further CO\textsubscript{2} injection or
seismic observations. During this forecasted period, the
CO\textsubscript{2} plume primarily ascends due to buoyancy, while a
portion (approximately 10\%) remains trapped in the pore spaces,
indicated in purple. This phenomenon, known as residual trapping (Rahman
et al. 2016), is a critical factor in assessing the long-term storage
capabilities of GCS projects. Initial forecasts tended to underestimate
the extent of CO\textsubscript{2} sequestration through residual
trapping. In contrast, simulations driven by the updated permeability
models not only provide a more accurate estimation of the permanently
stored CO\textsubscript{2} volume but also closely match the ground
truth CO2 plume's boundaries, even without any further monitoring data.

These findings underscore the potential of the permeability inversion
framework not only in refining current CO\textsubscript{2} plume
delineations but also in offering reliable predictions for future plume
behavior. Such predictive accuracy is invaluable for GCS project
managers and practitioners, enabling informed decision-making regarding
well control adjustments or intervention strategies to optimize
long-term CO\textsubscript{2} storage safety and efficiency, as
highlighted by recent studies in the field (Ringrose 2020).

\section{Limitations}\label{limitations}

While our case studies offer promising insights, it's crucial to
acknowledge the assumptions underpinning our approach and recognize the
inherent limitations that merit further investigation. Additionally, we
explore the potential for integrating this 4D processing workflow with
other reservoir characterization and management strategies.

\subsection{Reservoir simulation}\label{reservoir-simulation}

Our study assumes known values for all multiphase flow model parameters
except permeability with a known relationship between horizontal and
vertical permeability. Other parameters that include significant
rock-dependent variation include porosity, relative permeability
functions, residual water saturation, temperature and capillary pressure
were kept constant in the simulations to isolate the impact of
permeability on seismic data. In practice, a multi-parameter inversion,
potentially incorporating constraints like the Kozeny-Carman
relationship (Costa 2006), could offer a more holistic view by jointly
inverting for porosity and permeability. In addition, our assumption
that supercritical CO\textsubscript{2} miscibility in the resident brine
is low could be removed by considering a compositional flow model that
introduces additional uncertain parameters. The feasibility of such
approaches hinges on the availability of a differentiable reservoir
simulator, like
\href{https://github.com/sintefmath/JutulDarcy.jl}{JutulDarcy.jl}, or
the use of deep neural networks to approximate the physics of multiphase
flow (Grady et al. 2023; Louboutin, Yin, et al. 2023; Yin, Orozco, et
al. 2023b) and serve as a surrogate during inversion. Moreover, the
simplistic assumption of adherence to multiphase flow equations may not
hold in scenarios involving CO\textsubscript{2} leakage, necessitating
robust leakage detection methodologies (Erdinc et al. 2022; Yin, Erdinc,
et al. 2023).

\subsection{Rock physics}\label{rock-physics}

The case studies currently ignore the pressure effect on the wave
properties (MacBeth 2004; MacBeth, Floricich, and Soldo 2006). While
this can be justified for some GCS projects where the pressure change is
relatively subtle, the inversion framework can be extended to honor the
relationship between pressure and wave properties. The patchy saturation
model employed may also not fully capture the complexities of real-world
reservoirs, indicating a need for calibration of the rock physics model
against actual reservoir and seismic data.

\subsection{Wave physics}\label{wave-physics}

The omission of updates to the brine-filled baseline velocity model
represents a simplification that warrants further exploration. Future
research could extend the framework to jointly update this baseline
alongside permeability, incorporating additional parameters like shear
wavespeed and density. Quantifying uncertainties in velocity (Yin,
Orozco, et al. 2023a) and permeability models remains a critical
challenge for enhancing the reliability of inversion results.

\section{Discussion and conclusion}\label{discussion-and-conclusion}

This feasibility study, centered on the Compass model, elucidates the
inversion framework's capacity to refine permeability estimation,
offering high-resolution spatial insights that complement traditional
well measurements. Despite the limitations related to recovery
perfection and the reliance on initial models, the permeability updates
within the CO\textsubscript{2} plume region deliver actionable
intelligence for reservoir modelers, potentially reducing the cycle time
in 4D processing workflows.

Crucially, our study underscores the untapped potential of time-lapse
seismic data in reservoir permeability updates, bridging the gap between
seismic data and permeability through a novel end-to-end inversion
approach. While acknowledging the necessity for further validation
before field application, we posit that the integration of multiple
physical disciplines heralds a promising future for addressing the
complexities of time-lapse reservoir monitoring.

Embracing this multiphysics approach edges us closer to realizing a
digital twin for subsurface reservoir monitoring and management
(Herrmann 2023), highlighting the transformative potential of such
integrative methodologies in enhancing the precision and efficiency of
reservoir characterization and decision-making processes.

\section{Acknowledgement}\label{acknowledgement}

The authors gratefully acknowledge the contribution of OpenAI's ChatGPT
for refining sentence structure and enhancing the overall readability of
this manuscript. This research was carried out with the support of
Georgia Research Alliance and partners of the ML4Seismic Center. This
research was also supported in part by the US National Science
Foundation grant OAC 2203821.

\section{References}\label{references}

\phantomsection\label{refs}


\begin{CSLReferences}{1}{0}
\bibitem[\citeproctext]{ref-avseth2010quantitative}
Avseth, Per, Tapan Mukerji, and Gary Mavko. 2010. \emph{Quantitative
Seismic Interpretation: Applying Rock Physics Tools to Reduce
Interpretation Risk}. Cambridge university press.

\bibitem[\citeproctext]{ref-bloice2017}
Bloice, Marcus D., Christof Stocker, and Andreas Holzinger. 2017.
{``Augmentor: An Image Augmentation Library for Machine Learning.''}
\url{https://doi.org/10.48550/ARXIV.1708.04680}.

\bibitem[\citeproctext]{ref-bosch2007}
Bosch, Miguel, Luis Cara, Juan Rodrigues, Alonso Navarro, and Manuel
Díaz. 2007. {``A Monte Carlo Approach to the Joint Estimation of
Reservoir and Elastic Parameters from Seismic Amplitudes.''}
\emph{GEOPHYSICS} 72 (6): O29--39.
\url{https://doi.org/10.1190/1.2783766}.

\bibitem[\citeproctext]{ref-bosch2010}
Bosch, Miguel, Tapan Mukerji, and Ezequiel F. Gonzalez. 2010. {``Seismic
Inversion for Reservoir Properties Combining Statistical Rock Physics
and Geostatistics: A Review.''} \emph{GEOPHYSICS} 75 (5): 75A165--76.
\url{https://doi.org/10.1190/1.3478209}.

\bibitem[\citeproctext]{ref-chadwick2010quantitative}
Chadwick, Andy, Gareth Williams, Nicolas Delepine, Vincent Clochard,
Karine Labat, Susan Sturton, Maike-L Buddensiek, et al. 2010.
{``Quantitative Analysis of Time-Lapse Seismic Monitoring Data at the
Sleipner {CO2} Storage Operation.''} \emph{The Leading Edge} 29 (2):
170--77.

\bibitem[\citeproctext]{ref-chadwick2009}
Chadwick, R. A., D. Noy, R. Arts, and O. Eiken. 2009. {``Latest
Time-Lapse Seismic Data from Sleipner Yield New Insights into CO2 Plume
Development.''} \emph{Energy Procedia} 1 (1): 2103--10.
\url{https://doi.org/10.1016/j.egypro.2009.01.274}.

\bibitem[\citeproctext]{ref-costa2006permeability}
Costa, Antonio. 2006. {``Permeability-Porosity Relationship: A
Reexamination of the Kozeny-Carman Equation Based on a Fractal
Pore-Space Geometry Assumption.''} \emph{Geophysical Research Letters}
33 (2).

\bibitem[\citeproctext]{ref-e.jones2012}
E. Jones, C., J. A. Edgar, J. I. Selvage, and H. Crook. 2012.
{``Building Complex Synthetic Models to Evaluate Acquisition Geometries
and Velocity Inversion Technologies.''} \emph{Proceedings}, June.
\url{https://doi.org/10.3997/2214-4609.20148575}.

\bibitem[\citeproctext]{ref-eikrem2016}
Eikrem, Kjersti Solberg, Geir Nævdal, Morten Jakobsen, and Yan Chen.
2016. {``Bayesian Estimation of Reservoir Properties{\textemdash}effects
of Uncertainty Quantification of 4D Seismic Data.''} \emph{Computational
Geosciences} 20 (6): 1211--29.
\url{https://doi.org/10.1007/s10596-016-9585-0}.

\bibitem[\citeproctext]{ref-erdinc2022}
Erdinc, Huseyin Tuna, Abhinav Prakash Gahlot, Ziyi Yin, Mathias
Louboutin, and Felix J. Herrmann. 2022. {``De-Risking Carbon Capture and
Sequestration with Explainable CO2 Leakage Detection in Time-Lapse
Seismic Monitoring Images.''}
\url{https://doi.org/10.48550/ARXIV.2212.08596}.

\bibitem[\citeproctext]{ref-FURRE20173916}
Furre, Anne-Kari, Ola Eiken, Håvard Alnes, Jonas Nesland Vevatne, and
Anders Fredrik Kiær. 2017. {``20 Years of Monitoring {CO2}-Injection at
Sleipner.''} \emph{Energy Procedia} 114: 3916--26.
https://doi.org/\url{https://doi.org/10.1016/j.egypro.2017.03.1523}.

\bibitem[\citeproctext]{ref-gahlot2023inference}
Gahlot, Abhinav Prakash, Huseyin Tuna Erdinc, Rafael Orozco, Ziyi Yin,
and Felix J Herrmann. 2023. {``Inference of CO2 Flow Patterns--a
Feasibility Study.''} \emph{arXiv Preprint arXiv:2311.00290}.

\bibitem[\citeproctext]{ref-grady2023model}
Grady, Thomas J, Rishi Khan, Mathias Louboutin, Ziyi Yin, Philipp A
Witte, Ranveer Chandra, Russell J Hewett, and Felix J Herrmann. 2023.
{``Model-Parallel Fourier Neural Operators as Learned Surrogates for
Large-Scale Parametric PDEs.''} \emph{Computers \& Geosciences}, 105402.

\bibitem[\citeproctext]{ref-herrmann2023president}
Herrmann, Felix J. 2023. {``President's Page: Digital Twins in the Era
of Generative AI.''} \emph{The Leading Edge} 42 (11): 730--32.

\bibitem[\citeproctext]{ref-hicks2016}
Hicks, Erik, Henning Hoeber, Marianne Houbiers, Séverine Pannetier
Lescoffit, Andrew Ratcliffe, and Vetle Vinje. 2016. {``Time-Lapse
Full-Waveform Inversion as a Reservoir-Monitoring Tool {\textemdash} A
North Sea Case Study.''} \emph{The Leading Edge} 35 (10): 850--58.
\url{https://doi.org/10.1190/tle35100850.1}.

\bibitem[\citeproctext]{ref-hu2022}
Hu, Qi, Dario Grana, and Kristopher A Innanen. 2022. {``Feasibility of
Seismic Time-Lapse Monitoring of CO2 with Rock Physics Parametrized Full
Waveform Inversion.''} \emph{Geophysical Journal International} 233 (1):
402--19. \url{https://doi.org/10.1093/gji/ggac462}.

\bibitem[\citeproctext]{ref-krogstad2015mrst}
Krogstad, Stein, Knut--Andreas Lie, Olav Møyner, Halvor Møll Nilsen,
Xavier Raynaud, and Bård Skaflestad. 2015. {``MRST-AD--an Open-Source
Framework for Rapid Prototyping and Evaluation of Reservoir Simulation
Problems.''} In \emph{SPE Reservoir Simulation Conference?},
D022S002R004. SPE.

\bibitem[\citeproctext]{ref-li2020coupled}
Li, Dongzhuo, Kailai Xu, Jerry M Harris, and Eric Darve. 2020.
{``Coupled Time-Lapse Full-Waveform Inversion for Subsurface Flow
Problems Using Intrusive Automatic Differentiation.''} \emph{Water
Resources Research} 56 (8): e2019WR027032.

\bibitem[\citeproctext]{ref-louboutin2018dae}
Louboutin, Mathias, Fabio Luporini, Michael Lange, Navjot Kukreja,
Philipp A. Witte, Felix J. Herrmann, Paulius Velesko, and Gerard J.
Gorman. 2019. {``Devito (V3.1.0): An Embedded Domain-Specific Language
for Finite Differences and Geophysical Exploration.''}
\emph{Geoscientific Model Development}.
\url{https://doi.org/10.5194/gmd-12-1165-2019}.

\bibitem[\citeproctext]{ref-louboutin2023a}
Louboutin, Mathias, Philipp Witte, Ziyi Yin, Henryk Modzelewski, Kerim,
Grant Bruer, Carlos Da Costa, and Peterson Nogueira. 2023.
\emph{Slimgroup/JUDI.jl: V3.3.8}. Zenodo.
\url{https://doi.org/10.5281/ZENODO.8356884}.

\bibitem[\citeproctext]{ref-louboutin2023}
Louboutin, Mathias, Ziyi Yin, Rafael Orozco, Thomas J. Grady, Ali
Siahkoohi, Gabrio Rizzuti, Philipp A. Witte, Olav Møyner, Gerard J.
Gorman, and Felix J. Herrmann. 2023. {``Learned Multiphysics Inversion
with Differentiable Programming and Machine Learning.''} \emph{The
Leading Edge} 42 (7): 474--86.
\url{https://doi.org/10.1190/tle42070474.1}.

\bibitem[\citeproctext]{ref-lumley20104d}
Lumley, David. 2010. {``4D Seismic Monitoring of CO 2 Sequestration.''}
\emph{The Leading Edge} 29 (2): 150--55.

\bibitem[\citeproctext]{ref-lumley2001time}
Lumley, David E. 2001. {``Time-Lapse Seismic Reservoir Monitoring.''}
\emph{Geophysics} 66 (1): 50--53.

\bibitem[\citeproctext]{ref-luporini2020architecture}
Luporini, Fabio, Mathias Louboutin, Michael Lange, Navjot Kukreja,
Philipp Witte, Jan Hückelheim, Charles Yount, Paul HJ Kelly, Felix J
Herrmann, and Gerard J Gorman. 2020. {``Architecture and Performance of
Devito, a System for Automated Stencil Computation.''} \emph{ACM
Transactions on Mathematical Software (TOMS)} 46 (1): 1--28.

\bibitem[\citeproctext]{ref-macbeth2004}
MacBeth, Colin. 2004. {``A Classification for the Pressure{-}Sensitivity
Properties of a Sandstone Rock Frame.''} \emph{GEOPHYSICS} 69 (2):
497--510. \url{https://doi.org/10.1190/1.1707070}.

\bibitem[\citeproctext]{ref-macbeth2006}
MacBeth, Colin, Mariano Floricich, and Juan Soldo. 2006. {``Going
Quantitative with 4D Seismic Analysis.''} \emph{Geophysical Prospecting}
54 (3): 303--17. \url{https://doi.org/10.1111/j.1365-2478.2006.00536.x}.

\bibitem[\citeproctext]{ref-muxf8yner2023}
Møyner, Olav, and Grant Bruer. 2023. \emph{Sintefmath/JutulDarcy.jl:
V0.2.1}. Zenodo. \url{https://doi.org/10.5281/ZENODO.7775736}.

\bibitem[\citeproctext]{ref-pruess2011numerical}
Pruess, Karsten, and Jan Nordbotten. 2011. {``Numerical Simulation
Studies of the Long-Term Evolution of a {CO2} Plume in a Saline Aquifer
with a Sloping Caprock.''} \emph{Transport in Porous Media} 90 (1):
135--51.

\bibitem[\citeproctext]{ref-rahman2016}
Rahman, Taufiq, Maxim Lebedev, Ahmed Barifcani, and Stefan Iglauer.
2016. {``Residual Trapping of Supercritical CO2 in Oil-Wet Sandstone.''}
\emph{Journal of Colloid and Interface Science} 469 (May): 63--68.
\url{https://doi.org/10.1016/j.jcis.2016.02.020}.

\bibitem[\citeproctext]{ref-rasmussen2021open}
Rasmussen, Atgeirr Flø, Tor Harald Sandve, Kai Bao, Andreas Lauser,
Joakim Hove, Bård Skaflestad, Robert Klöfkorn, et al. 2021. {``The Open
Porous Media Flow Reservoir Simulator.''} \emph{Computers \& Mathematics
with Applications} 81: 159--85.

\bibitem[\citeproctext]{ref-ringrose2020store}
Ringrose, Philip. 2020. \emph{How to Store {CO2} Underground: Insights
from Early-Mover {CCS} Projects}. Springer.

\bibitem[\citeproctext]{ref-settgast2018geosx}
Settgast, Randolph R, JA White, BC Corbett, A Vargas, C Sherman, P Fu, C
Annavarapu, et al. 2018. {``Geosx Simulation Framework.''} Lawrence
Livermore National Lab.(LLNL), Livermore, CA (United States).

\bibitem[\citeproctext]{ref-stacey2017validation}
Stacey, Robert W, and Michael J Williams. 2017. {``Validation of ECLIPSE
Reservoir Simulator for Geothermal Problems.''} \emph{GRC Transactions}
41: 2095--2109.

\bibitem[\citeproctext]{ref-tarantola1984inversion}
Tarantola, Albert. 1984. {``Inversion of Seismic Reflection Data in the
Acoustic Approximation.''} \emph{Geophysics} 49 (8): 1259--66.

\bibitem[\citeproctext]{ref-vasco2008}
Vasco, D. W., Henk Keers, Jalal Khazanehdari, and Anthony Cooke. 2008.
{``Seismic Imaging of Reservoir Flow Properties: Resolving Water Influx
and Reservoir Permeability.''} \emph{GEOPHYSICS} 73 (1): O1--13.
\url{https://doi.org/10.1190/1.2789395}.

\bibitem[\citeproctext]{ref-vasco2004}
Vasco, Don W., Akhil Datta-Gupta, Ron Behrens, Pat Condon, and James
Rickett. 2004. {``Seismic Imaging of Reservoir Flow Properties:
Time{-}Lapse Amplitude Changes.''} \emph{GEOPHYSICS} 69 (6): 1425--42.
\url{https://doi.org/10.1190/1.1836817}.

\bibitem[\citeproctext]{ref-wei2017}
Wei, Lei, Prasenjit Roy, Todd Dygert, David Grimes, and Mason Edwards.
2017. {``Estimation of Reservoir Pressure and Saturation Changes from 4D
Inverted Elastic Properties.''} \emph{SEG Technical Program Expanded
Abstracts 2017}, August.
\url{https://doi.org/10.1190/segam2017-17733336.1}.

\bibitem[\citeproctext]{ref-witte2018alf}
Witte, Philipp A., Mathias Louboutin, Navjot Kukreja, Fabio Luporini,
Michael Lange, Gerard J. Gorman, and Felix J. Herrmann. 2019. {``A
Large-Scale Framework for Symbolic Implementations of Seismic Inversion
Algorithms in Julia.''} \emph{Geophysics} 84 (3): F57--71.
\url{https://doi.org/10.1190/geo2018-0174.1}.

\bibitem[\citeproctext]{ref-yin2023derisking}
Yin, Ziyi, Huseyin Tuna Erdinc, Abhinav Prakash Gahlot, Mathias
Louboutin, and Felix J Herrmann. 2023. {``Derisking Geologic Carbon
Storage from High-Resolution Time-Lapse Seismic to Explainable Leakage
Detection.''} \emph{The Leading Edge} 42 (1): 69--76.

\bibitem[\citeproctext]{ref-yin2023wise}
Yin, Ziyi, Rafael Orozco, Mathias Louboutin, and Felix J Herrmann.
2023a. {``WISE: Full-Waveform Variational Inference via Subsurface
Extensions.''} \emph{arXiv Preprint arXiv:2401.06230}.

\bibitem[\citeproctext]{ref-yin2023}
Yin, Ziyi, Rafael Orozco, Mathias Louboutin, and Felix J. Herrmann.
2023b. {``Solving Multiphysics-Based Inverse Problems with Learned
Surrogates and Constraints.''} \emph{Advanced Modeling and Simulation in
Engineering Sciences} 10 (1).
\url{https://doi.org/10.1186/s40323-023-00252-0}.

\bibitem[\citeproctext]{ref-yin2022SEGlci}
Yin, Ziyi, Ali Siahkoohi, Mathias Louboutin, and Felix J. Herrmann.
2022. {``Learned Coupled Inversion for Carbon Sequestration Monitoring
and Forecasting with Fourier Neural Operators.''}
\url{https://doi.org/10.1190/image2022-3722848.1}.

\bibitem[\citeproctext]{ref-zhang2014}
Zhang, Zhishuai, Behnam Jafarpour, and Lianlin Li. 2014. {``Inference of
Permeability Heterogeneity from Joint Inversion of Transient Flow and
Temperature Data.''} \emph{Water Resources Research} 50 (6): 4710--25.
\url{https://doi.org/10.1002/2013wr013801}.

\bibitem[\citeproctext]{ref-ziyiyin2023}
Ziyi Yin, Grant Bruer, and Mathias Louboutin. 2023.
\emph{Slimgroup/JutulDarcyRules.jl: V0.2.6}. Zenodo.
\url{https://doi.org/10.5281/ZENODO.8172164}.

\end{CSLReferences}
\end{document}